\documentclass[a4paper, 11pt]{article}
\usepackage{comment} 
\usepackage{fullpage} 
\usepackage{graphicx}
\usepackage{cite}
\usepackage{bm}
\usepackage{float}
\usepackage{amsmath}
\usepackage[utf8]{inputenc}
\usepackage{amssymb}
\usepackage{mathrsfs}
\usepackage{wrapfig}
\usepackage{enumitem}
\usepackage[english]{babel}
\usepackage{comment}
\usepackage{bbold}  
\usepackage{hyperref}
\usepackage[usenames,dvipsnames]{xcolor} 
\usepackage{xcolor}
\usepackage{setspace}
\usepackage{abstract}
\usepackage{sectsty}
\sectionfont{\large}
\usepackage{tablefootnote}

\newcommand\myfontsize{\fontsize{10pt}{12pt}\selectfont}
\usepackage[T1]{fontenc}
\usepackage[margin=0.94in]{geometry}
\usepackage[nointegrals]{wasysym}

\begin{document}
\null\hfill LMU-ASC 34/23\; \\
\null\hfill MPP-2023-256\; \\
\vspace*{\fill}
\begin{center}
    \LARGE\textbf{\textsc{ \textcolor{Black}{On the Degrees of Freedom of $R^2$ Gravity\\ in Flat Spacetime} }}    
    
     \normalsize\textsc{Anamaria Hell,$^{1}$ Dieter Lüst,$^{2,\;3}$ George Zoupanos$^{3,\; 4,\; 5, \; 6 }$}
\end{center}

\begin{center}
    $^{1}$ \textit{Kavli IPMU (WPI), UTIAS,\\ The University of Tokyo,\\ Kashiwa, Chiba 277-8583, Japan}\\
    $^{2}$\textit{Arnold Sommerfeld Center for Theoretical Physics,\\
Ludwig–Maximilians–Universität München\\
Theresienstraße 37, 80333 Munich, Germany}\\
$^{3}$\textit{Max–Planck–Institut für Physik (Werner–Heisenberg–Institut)\\
Föhringer Ring 6, 80805 Munich, Germany}\\ 
$^{4}$\textit{Physics Department, National Technical University,\\ Zografou campus
157 80 Athens, Greece}\\
$^{5}$\textit{Theory Department, CERN,\\ Geneva 1211, Switzerland
}\\
$^{6}$\textit{Institut für Theoretische Physik der Universität Heidelberg,\\Philosophenweg 16, 69120 Heidelberg, Germany}
\end{center}
\renewcommand{\abstractname}{\textsc{\textcolor{Black}{Abstract}}}

\begin{abstract}
     We study the degrees of freedom of $R^2$ gravity in flat spacetime with two approaches. By rewriting the theory a la Stueckelberg, and implementing Lorentz-like gauges to the metric perturbations, we confirm that the pure theory propagates one scalar degree of freedom, while the full theory contains two tensor modes in addition. We then consider the degrees of freedom by directly examining the metric perturbations. We show that the degrees of freedom of the full theory match with those obtained with the manifestly covariant approach. In contrast, we find that the pure $R^2$ gravity has no degrees of freedom. We show that a similar discrepancy between the two approaches appears also in a theory dual to the three-form, and appears due to the Lorentz-like gauges, which lead to the fictitious modes even after the residual gauge redundancy has been taken into account. At first sight, this implies a discontinuity between the full theory and the pure case. By studying the first-order corrections of the full $R^2$ gravity beyond the linear regime, we show that at high-energies, both scalar and tensor degrees of freedom become strongly coupled. This implies that the apparent discontinuity of pure and full $R^2$ gravity is just an artefact of the perturbation theory, and further supports the absence of degrees of freedom in the pure $R^2$ gravity. 
\end{abstract}

\thispagestyle{empty} 
\vspace*{\fill}
\clearpage
\pagenumbering{arabic} 
\newpage


\renewcommand*\contentsname{\Large {\textsc{Content}}}
\tableofcontents

\section{\Large\textsc{Introduction}}
It has been more than a century since the introduction of Einstein's General Relativity (GR). To date, this theory, with action made out of the Ricci scalar, has passed numerous experimental tests. Yet, theoretical principles and recent cosmological observations challenge it, bringing a significant interest in its extensions and alternative theories of gravity \cite{CANTATA:2021ktz}.

Among these, \textsc{Quadratic Gravity (QG)} -- a theory which extends the Einstein-Hilbert action by including also quadratic powers of the Ricci scalar and Weyl tensor -- is particularly interesting \cite{Stelle:1976gc, Stelle:1977ry, Boulware:1983td, David:1984uv, Deser:2007vs, Buchbinder:1987vp, tHooft:2011aa, Alvarez-Gaume:2015rwa, Utiyama:1962sn, Horowitz:1984wv, Park:2012ds, Lu:2011ks, DeFelice:2023apt, Salvio:2018crh, DeFelice:2023kpl, Tadros:2023teq, Buoninfante:2023ryt, Konitopoulos:2023wst, Manolakos:2019fle, Manolakos:2021rcl, Tekin:2016vli}. As initially noticed in \cite{Utiyama:1962sn} and later shown in \cite{Stelle:1976gc, Stelle:1977ry}, QG improves the renormalizability of Einstein's GR and, is argued to be asymptotically free\footnote{See also \cite{Capper:1975ig}, for an argument against these claims. \hfill} \cite{Stelle:1976gc, Fradkin:1981iu, Julve:1978xn}. 
Yet, the full QG is not unitary -- it contains Ostrogradsky ghosts that make its energy unbounded from below \cite{Ostrogradsky:1850fid, Riegert:1984hf}. Notably, in the case of conformal gravity, which has only the Weyl tensor sqared, it was shown that these ghosts could be removed by imposition of the boundary conditions in the de Sitter and Minkowski space \cite{Maldacena:2011mk, Hell:2023rbf, Anastasiou:2016jix, Anastasiou:2020mik}. 

\textsc{$R^2$ Gravity} -- a theory that involves only the terms linear and quadratic in Ricci scalar -- is the simplest case of $f(R)$ gravity, and the only combination of QG that is ghost-free \cite{Sotiriou:2008rp}.\footnote{See also \cite{Casado-Turrion:2023rni}, for recent claims on the non-viability of a class of $f(R)$ theories. } This model, also known as the Starobinsky inflation, is one of the most promising inflationary models to date \cite{Mukhanov:1989rq, Mukhanov1987, Starobinsky:1979ty, Mukhanov:1981xt, Guth:1980zm, Linde:1981mu, Albrecht:1982wi, Planck:2018jri}. It has also been extensively studied in the context of supergravity \cite{Cecotti, Ellis:2013xoa, Farakos:2013cqa, Ellis:2013nxa, Kallosh:2013lkr, Ferrara:2013wka, Ferrara:2013pla, Ellis:2014cma, Ferrara:2014ima, Ferrara:2014fqa, Ferrara:2014yna, Ferrara:2014cca, Dalianis:2014aya, Diamandis:2014vxa, Lahanas:2015jwa, Kounnas:2014gda, Farakos:2017mwd} and black holes \cite{JAcobson, Frolov:2009qu, Nelson:2010ig, Lu:2015cqa}. 

\textsc{The Pure $R^2$ Gravity} is a special case of $R^2$ gravity, which contains only the square of the Ricci Scalar. It is scale-invariant and, in addition, has restricted Weyl symmetry, in which the Weyl transformation parameter satisfies the wave equation \cite{Oda:2020wdd, Edery:2014nha, Edery:2015wha}. To date, many classical solutions of this theory have been investigated \cite{Kehagias:2015ata, Pravda:2016fue, Podolsky:2018pfe, Gurses:2012db, Azreg-Ainou:2023qtf, Nguyen:2023ufi, Nguyen:2023kwr, Nguyen:2022blj, Dent:2016efw, Duplessis:2015xva, Edery:2018jyp, Perapechka:2018iqo, Bahamonde:2018zcq}. Interestingly, in the Newtonian limit, the potential of this theory has only a confining part \cite{Alvarez-Gaume:2015rwa}.  

As shown in \cite{Whitt:1984pd, Mukhanov:1989rq}, $R^2$ gravity is conformally equivalent to Einstein gravity coupled to the scalar field. However, for flat space-time, the transformation to the Einstein frame is singular in the pure case \cite{Alvarez-Gaume:2015rwa}. To study the \textit{degrees of freedom (dof)}, the authors of \cite{Alvarez-Gaume:2015rwa} have thus considered perturbations around the Minkowski metric for the original action. They have found that in the case of the full $R^2$ gravity, the theory has a massive scalar mode and two tensor modes that describe gravitational waves. In contrast, in the pure case, they have shown that the tensor modes are absent, and the theory describes only a single massless scalar mode. To arrive at this result, their analysis contained two key ingredients -- the Stueckelberg trick and the Lorentz-like gauges. 

\textsc{The Stueckelberg Trick} is a way to introduce additional \textit{dof} to the theory in a manifestly covariant way, such that the resulting theory has gauge redundancy \cite{Ruegg:2003ps, Stueckelberg}. Originally, it was introduced in Proca theory, the theory of a massive vector field which describes three degrees of freedom -- a longitudinal scalar mode, and two transverse vector ones \cite{Proca}. In contrast to Maxwell's electrodynamics, this theory has no gauge redundancy. However, by rewriting the original vector field as a sum of a new one with a 4-derivative of a scalar field, the gauge redundancy appears for the new variables. This trick has also been extended to other theories, such as massive Yang-Mills theory, with mass added \textit{by hand}, and massive gravity \cite{deRham:2014zqa, Hinterbichler:2011tt, Kunimasa, Vainshtein, Gambuti:2020onb, Huang:2007xf, Dvali:2007kt, deRham:2011qq}. 

In the case of gravity, by rewriting the metric perturbations \textit{a la Stueckelberg}, one adds scalar and vector \textit{dof} to the theory, thus introducing gauge redundancy. By employing \textsc{the Lorentz-Like Gauges} -- conditions that the 4-divergence of the tensor and vector components is vanishing -- the resulting relation between old and new variables resembles the helicity decomposition.\footnote{Although, as pointed out in \cite{Hinterbichler:2011tt}, the Stueckelberg trick is \textbf{not} a decomposition, but an introduction of new variables.\hfill} This thus makes the Stueckelberg trick favorable to study the \textit{dof} of theories that even initially have gauge redundancy (see \cite{Alvarez:2018lrg, Buoninfante:2023ryt, Ghosh:2023gvc, Edery:2023hxl, Kubo:2022jwu, Kubo:2022dlx, Kubo:2022lja, Kamimura:2021wzf, Dalianis:2020nuf, Hinterbichler:2015soa} for studies including QG). 

While this procedure is favored, as it preserves the manifest Lorentz covariance, the resulting action can appear to have ghost-like \textit{dof}.  This was the case in \cite{Alvarez-Gaume:2015rwa} for the scalar modes, and appears even for linearized gravity alone. By noting that the Stueckelberg trick introduces additional gauge redundancy on top of the invariance under the infinitesimal coordinate transformations, it can be shown that all ghost-like \textit{dof} are not physical modes of the theory. Nevertheless, if one wishes to study the nature of these degrees of freedom, or quantize the theory, a more convenient approach would be to express the action such that it only contains physical modes.  Thus, the goal of this paper is to answer -- \textit{Is there an alternative way to describe the $R^2$ gravity in flat space, in which the action is expressed only in terms of the propagating \textit{dof}?}

We will see that this is possible with a direct approach, based on the  \textsc{Cosmological Perturbation Theory (CPT)} -- a theory that brought upon our understanding of the origin of galaxies and the large-scale structure \cite{Chibisov:1982nx, Mukhanov:1981xt, Mukhanov:1990me, Sasaki:1986hm, Kodama:1984ziu}. The core of CPT lies in the study of the evolution of the metric perturbations in an expanding Universe, by decomposing them according to the spatial rotations. 
As a result, the theory contains scalar, vector, and tensor modes that can be separately studied. This procedure can be even applied to vector theories, and theories of 2-form and 3-form fields \cite{Chamseddine:2012gh, Hell:2021wzm}. Notably, with the tools of CPT, it was possible to resolve one of the contemporary theoretical puzzles -- the apparent discontinuity of the massless limit of massive Yang-Mills theory, with mass added \textit{by hand} --  and show that the limit is smooth, in contrast to the previous approaches \cite{Hell:2021oea}. 

In this paper, we will use the tools of CPT to study the \textit{dof} of the full and pure $R^2$ gravity in the flat space. Surprisingly, we will find that while the manifestly covariant and direct approach gives the same number of degrees of freedom for the full theory, this is not the case for the pure $R^2$ gravity -- in the flat background, we will show that this theory has no \textit{dof}.  
Thus we will find that the scalar sector of $R^2$ gravity has a striking resemblance to the theory of a 3-form, which if massless has no \textit{dof}, while in the massive case describes one massive pseudoscalar \cite{ Kawai, Trugenberger, Quevedo, 2001Smailagic, 2002Casini, Auria, Buchbinder, Dalmazi, Shifman, Garcia, Kuzenko, CF1980, C2019, Dvali}. Moreover, by studying the theories two-ways, via the manifestly covariant approach \cite{Alvarez-Gaume:2015rwa} and a direct one, we will see that the discrepancy in the \textit{dof} that appears for $R^2$ gravity persists also for the 3-form theory, if rewritten in terms of its dual vector field, through the Levi-Civita symbol. 

Curiously, the mismatch between the \textit{dof}of the full and pure $R^2$ theory implies the presence of a discontinuity. \textsc{The vDVZ Discontinuity} -- discrepancy between the predictions of massive Fiersz-Pauli theory and linearised GR for the deflection of starlight and the precession of the perihelia of Mercury -- seems, at first sight, to be so significant that one could exclude massive gravity as a possible theory of nature \cite{Fierz:1939ix, vanDam:1970vg, Zakharov:1970cc}. A similar discontinuity appears even in the massive Yang-Mills theory, in the form of the corrections to the propagator of the gauge bosons at one loop, if their mass is added \textit{by hand}, and treated by conventional perturbative methods \cite{vanDam:1970vg}. At higher order corrections, it even manifests by the singular behavior of the perturbative expansion \cite{Veltman:1968ki, Reiff:1969pq, Slavnov:1972qb, Wong:1971er, Boulware:1970zc}.  Moreover, the apparent discontinuity of the massless limit also arises in other gauge theories, such as massive two and three forms, or even in Proca theory, if their self-interactions are taken into account, and manifests in the apparent singular behavior of the perturbative series in the massless limit \cite{Hell:2021wzm}. 

The origin of the discrepancy between the behavior of the massive and massless gauge theories lies in the modes that are absent in the massless theory. \textsc{The Vainshtein Mechanism}, however, resolves this pathology. As originally pointed out in \cite{Vainshtein:1972sx}, the longitudinal mode that introduces the vDVZ pathology in massive gravity becomes strongly coupled due to the non-linear terms, and as a result decouples from the remaining degrees of freedom \cite{Deffayet:2001uk, Gruzinov:2001hp}. As a result, the apparent discontinuity in the massless limit is just an artefact of the perturbation theory, and the results of GR become restored beyond this scale. Moreover, the same mechanism was applied for the massive Yang-Mills and self-interacting theories of Proca, Kalb-Ramond, and 3-form fields\footnote{See \cite{Hu:2023juh, Hu:2023xcf} for an alternative approach to the strong coupling problem in $f(T)$ gravity.} \cite{Hell:2021wzm, Hell:2021oea, Dvali:2007kt}. 

As we will see, while the full $R^2$ gravity propagates a scalar mode together with two tensor ones, the pure case has no dof. Thus, at the linearized level, there might be a discontinuity between the two theories, in the limit when the parameter in front of the Einstein's contribution vanishes. However, by studying the first-order corrections of the full theory beyond the linear regime, we will show that this is just an artefact of the perturbation theory. In particular, we will find that at high energies, when the $R^2$ term dominates, both scalar and tensor modes become strongly coupled, showing that a mechanism similar to the case of massive gauge theories takes place.

The paper is organized as follows. First, we will study the two approaches to $R^2$ gravity -- the\textit{ manifestly covariant approach}, in which we will review the analysis of \cite{Alvarez-Gaume:2015rwa} and confirm its results, and the \textit{direct approach}, which uses the gauge-invariant variables, and gives rise to a disagreement in the \textit{dof}for the pure case. We will then generalize this analysis to the theory dual to the 3-form, and show that the same discrepancy appears there as well. Once we identify the origin of the disagreement between the two approaches, we will study the high.energy limit of $R^2$ gravity, and the behavior of its modes in the presence of the non-linear terms. 

\section{\Large\textsc{The Two Faces of $R^2$ Gravity}}

In this section, we will study the \textit{degrees of freedom (dof)} of the linearized $R^2$ gravity in the flat spacetime with two approaches -- the manifestly covariant one, in which we will closely follow the analysis of \cite{Alvarez-Gaume:2015rwa}, and a direct approach, that relies on the gauge-invariant metric perturbations. 

The \textit{full} $R^2$ gravity is described by the action 
\begin{equation}\label{StarobinskyAction}
    S_{full}=\int d^4x\sqrt{-g}\left[\frac{M^2}{2}R+\beta R^2\right]. 
\end{equation}
In addition to this theory, we will consider the pure $R^2$ gravity which is a special case of the above theory with action given by: 
\begin{equation}\label{pureR2}
    S_{pure}=\beta\int d^4x\sqrt{-g} R^2. 
\end{equation}

In order to study the \textit{dof}, let us first consider small perturbations of the metric around a flat background: 
\begin{equation}
    g_{\mu\nu}=\eta_{\mu\nu}+h_{\mu\nu}. 
\end{equation}
Then, the Lagrangian density corresponding to the action (\ref{StarobinskyAction}) up to quadratic terms in the metric perturbations is given by: 
\begin{equation}\label{Lagrangianfull}
  \begin{split}
        \mathcal{L}_{full}=\frac{M^2}{2}\left(2\partial_{\mu}h^{\mu\nu}\partial^{\alpha}h_{\alpha\nu}-\partial_{\alpha}h_{\mu\nu}\partial^{\alpha}h^{\mu\nu}-2\partial_{\mu}h^{\mu\nu}\partial_{\nu}h+\partial_{\mu}h\partial^{\mu}h\right)+\beta\left(\partial_{\mu}\partial_{\nu}h^{\mu\nu}-\Box h\right)^2,
  \end{split}
\end{equation}

where $h=h^{\mu}_{\mu}$. We can notice that the above expression involves higher derivatives of the metric perturbations. So the most natural question would be to investigate if the theory contains \textit{dof}that are ghosts. However, as we will see, this will not be the case. In order to show this, let us first study the theory with the manifestly covariant approach. 

\subsection{\large\textsc{Approaching the $R^2$ gravity \textit{a la Stueckelberg}}}

Following \cite{Alvarez-Gaume:2015rwa}, let us now apply the Stueckelberg trick to the metric perturbations, by rewriting the original metric perturbations as:
\begin{equation}\label{decompositionMCOV}
h_{\mu\nu}=l_{\mu\nu}^T+\partial_{\mu}A_{\nu}^T+\partial_{\nu}A_{\mu}^T+\left(\partial_{\mu}\partial_{\nu}-\frac{1}{4}\Box\eta_{\mu\nu}\right)\mu+\frac{1}{4}\lambda\eta_{\mu\nu}.
\end{equation}
We will impose the Lorentz-like conditions on the new vector and tensor fields: 
\begin{equation}\label{MCOVproperties1}
    \partial_{\mu}l^{T\mu}_{\nu}=0,\qquad \text{and}\qquad \partial_{\mu}A^{T\mu}=0,
\end{equation}
and, in addition, require:
\begin{equation}\label{MCOVproperties2}
    l^{T\mu}_{\mu}=0
\end{equation}
In this case, (\ref{Lagrangianfull}) becomes: 
\begin{equation}
    \mathcal{L}_{full}=\frac{M^2}{8}\left[-\partial_{\alpha}l_{\mu\nu}^T\partial^{\alpha}l^{T\mu\nu}+\frac{3}{8}\partial_{\alpha}\left(\Box\mu-\lambda\right)\partial^{\alpha}\left(\Box\mu-\lambda\right)\right]+\frac{9\beta}{19}\left(\Box\mu-\lambda\right)^2.
\end{equation}
Thus, we can see that this theory describes tensor and scalar modes, while the vector ones have dropped out from the action. However, the above fields are frame-dependent. Under infinitesimal coordinate transformations:
\begin{equation}\label{cooinv}
    x^{\mu}\to\Tilde{x}^{\mu}=x^{\mu}+\xi^{\mu}. 
\end{equation}
the above fields transform as: 
\begin{equation}
    \lambda\to\Tilde{\lambda}=\lambda-2\Box\kappa\qquad\text{and}\qquad \mu\to\Tilde{\mu}=\mu-2\kappa,
\end{equation}
where we have decomposed the parameter $\xi^{\mu}$ as 
\begin{equation}
    \xi^{\mu}=\xi^{T\mu}+\partial^{\mu}\kappa, \qquad \text{where}\qquad \partial_{\mu}\xi^{T\mu}=0. 
\end{equation}
We can form an invariant quantity: 
\begin{equation}
    \sigma=\lambda-\Box\mu,
\end{equation}
which remains unchanged with the transformation (\ref{cooinv}). With it, the Lagrangian density becomes: 
\begin{equation}\label{lagrangianNCOVgdof}
    \mathcal{L}_{pure}=-\frac{M^2}{8}\partial_{\alpha}l_{\mu\nu}^T\partial^{\alpha}l^{T\mu\nu}+\frac{9\beta}{16}\Box\sigma\left(\Box-m_{\sigma}^2\right)\sigma,
\end{equation}
in agreement with \cite{Alvarez-Gaume:2015rwa}, where
\begin{equation}
    m_{\sigma}^2=\frac{M^2}{12\beta}. 
\end{equation}
Let's now first consider the full $R^2$ gravity. We can notice that the second term of (\ref{lagrangianNCOVgdof}) indicates that the theory is plagued by ghosts -- it would describe a massive healthy \textit{dof} as well as the massless ghost one. However, this comes with a subtlety. The relation among the old metric perturbations and new ones, given in (\ref{decompositionMCOV}), is not a decomposition, but rather an introduction of the additional vector and scalar degrees of freedom \cite{Hinterbichler:2011tt}. 
Thus, in addition to the invariance under infinitesimal coordinate transformations (\ref{cooinv}), by performing such a trick we have introduced additional gauge invariance on top of the existing one, due to the presence of the new vector and scalar fields. We have already partially fixed the gauge to set the conditions (\ref{MCOVproperties1}) and (\ref{MCOVproperties2}). However, similarly to the Lorentz gauge-fixing in electrodynamics, this leaves us with residual gauge freedom \cite{Alvarez-Gaume:2015rwa}: 
\begin{equation}\label{resgaugeMCOV}
    l^T_{\mu\nu}\to \Tilde{l}_{\mu\nu}^T+\partial_{\mu}v_{\nu}+\partial_{\nu}v_{\mu}\qquad\qquad \sigma\to\Tilde{\sigma}=\sigma+2\partial^{\mu}s_{\mu},
\end{equation}
where 
\begin{equation}
    \partial^{\mu}v_{\nu}=0,\qquad \Box v_{\mu}=0\qquad \text{and}\qquad \partial_{\mu} s_{\nu}+\partial_{\nu} s_{\mu}=\frac{1}{2}\eta_{\mu\nu} \partial^{\gamma} s_{\gamma}. 
\end{equation}
By using it, we can remove one ghost-like \textit{dof}. Then, all together, the Lagrangian density (\ref{lagrangianNCOVgdof}) describes two tensor degrees of freedom and one healthy scalar mode. 

Let us now consider only the pure  $R^2$ gravity. In this case, the Lagrangian density becomes: 
\begin{equation}\label{lagrangianNCOVgdofpure}
    \mathcal{L}_{pure}=\frac{9\beta}{16}\Box\sigma\Box\sigma.
\end{equation}
In contrast to the full $R^2$ gravity, we can see that the tensor modes have disappeared, and the theory only describes scalar \textit{dof}, in agreement with \cite{Alvarez-Gaume:2015rwa}. We should note that even though it appears that another ghost-like \textit{dof} is contained in the theory, it can be removed in the same manner as in the previous case.

\subsection{\large\textsc{The Direct Approach }}

In the previous subsection we have investigated the degrees of freedom of the $R^2$ gravity in the flat space using the Stueckelberg trick together with the Lorentz-like conditions on the metric perturbations. We have seen that the pure $R^2$ gravity describes a single, massless, scalar \textit{dof}. By adding to this theory an Einstein-Hilbert term, this scalar becomes massive, and the theory in addition propagates two tensor modes that describe the gravitational waves.

However, while the previous approach keeps the Lorentz covariance manifest, it is not convenient. We have seen that the scalars appear with higher-time derivatives, and thus suggest a presence of ghosts. While these modes are not physical and can be removed with the residual gauge redundancy, it would be interesting to see the form of the action which contains only the physical \textit{dof}. We will explore such formulation of the theory in this section. In contrast to the previous approach, we will follow an alternative, direct one, that originates from the cosmological perturbations theory. 

\subsubsection{\textsc{The Pure $R^2$ Gravity}} 

Let's begin our analysis with the pure $R^2$ gravity. As in the previous section, we will decompose the metric in terms of the flat space-time background, and a small perturbation: 
\begin{equation}
    g_{\mu\nu}=\eta_{\mu\nu}+h_{\mu\nu}. 
\end{equation}
In this case, the Lagrangian density corresponding to the action (\ref{pureR2}), is given by: 
\begin{equation}\label{Lagrangianpure}
    \begin{split}
        \mathcal{L}_{pure}=&\beta\left(\partial_{\mu}\partial_{\nu}h^{\mu\nu}-\Box h\right)^2.
    \end{split}
\end{equation} 
In order to know the full spectrum of the theory, let us decompose the metric perturbations in the scalar, vector and tensor modes \cite{Cosmo}:
\begin{equation}\label{decompositionCPT}
    \begin{split}
        &h_{00}=2\phi\\
        &h_{0i}=B_{,i}+S_i,\qquad\qquad S_{i,i}=0\\
        &h_{ij}=2\psi\delta_{ij}+2E_{,ij}+F_{i,j}+F_{j,i}+h_{ij}^{T},\qquad\qquad F_{i,i}=0,\quad h_{ij,i}^{T}=0,\quad h_{ii}^{T}=0,
    \end{split}
\end{equation}
where $,i=\frac{\partial}{\partial x^i}$. Then, the above Lagrangian density becomes: 
\begin{equation}\label{LpureCPT1}
    \begin{split}
\mathcal{L}_{pure}=&\beta\left[36\Ddot{\psi}\Ddot{\psi}-48\Ddot{\psi}\Delta\psi+16\Delta\psi\Delta\psi \right.\\ & \left. +4\Delta\left(\phi+\Ddot{E}-\dot B\right)\left(6\Ddot{\psi}-4\Delta\psi+\Delta\left(\phi+\Ddot{E}-\dot B\right)\right)\right].
    \end{split}
\end{equation}
Here, $\dot{}$ denotes a derivative with respect to time. We can notice that as in the initial approach, both tensor and vector perturbations have dropped out from the action. Moreover, the scalars that are involved in this Lagrangian density depend on the choice of the coordinate frame. Upon the infinitesimal coordinate transformation: 
\begin{equation}
    x^{\mu}\to\Tilde{x}^{\mu}=x^{\mu}+\xi^{\mu},
\end{equation}
one can easily show that the scalar perturbations transform as: 
\begin{equation}
    \begin{split}
        \phi\to\Tilde{\phi}=\phi-\dot{\xi}_0\qquad \psi\to\Tilde{\psi}=\psi \qquad E\to\Tilde{E}=E-\zeta\qquad B\to\Tilde{B}=B-\xi_0-\dot{\zeta}.
    \end{split}
\end{equation}
Here, we have decomposed the infinitesimal parameter $\xi^{\mu}$ as
\begin{equation}
    (\xi_0, \xi_i),\qquad \text{and}\qquad \xi_i=\xi_i^T+\zeta_{,i}.
\end{equation}
Even though they are absent at the moment, let us note for completeness that the vector perturbations transform as: 
\begin{equation}
    S_i\to\Tilde{S}_i=S_i-\dot{\xi}_i^T\qquad\qquad F_i\to\Tilde{F}_i=F_i-\xi_i^T
\end{equation}
while the tensor modes are left unchanged: 
\begin{equation}
    h_{ij}^T\to\Tilde{h}_{ij}^T=h_{ij}^T. 
\end{equation}
 By defining the following \textit{gauge-invariant} quantities:
\begin{equation}
    \Phi=\phi-\left(\dot{B}-\Ddot{E}\right)\qquad \text{and}\qquad \Psi=\psi
\end{equation}
 known in the cosmological perturbation theory as \textit{the Bardeen potentials} \cite{Bardeen:1980kt}, the Lagrangian density (\ref{LpureCPT1}) becomes 
\begin{equation}\label{LpureCPT2}
    \begin{split}
\mathcal{L}_{pure}=4\beta\left[\Delta\Phi\Delta\Phi+2\Phi\Delta\left(3\Ddot{\Psi}-2\Delta\Psi\right)+9\Ddot{\Psi}\Ddot{\Psi}+4\Delta\Psi\Delta\Psi-12\Ddot{\Psi}\Delta\Psi\right].
    \end{split}
\end{equation}
This expression can be further simplified by defining the following \textit{gauge-invariant} quantity: 
\begin{equation}
    \chi=\Phi+\Psi.
\end{equation}
With it, (\ref{LpureCPT2}) becomes: 
\begin{equation}\label{LpureCPT3}
    \begin{split}
\mathcal{L}_{pure}&=4\beta\left[\Delta\chi\Delta\chi-6\Delta\chi\Box\Psi+9\Box\Psi\Box\Psi\right].
    \end{split}
\end{equation}
Clearly, the quadratic term for $\Psi$ appears with higher-order time derivatives. Naively, this could leave an impression that the theory does contain scalar ghosts. However, in contrast to the previous approach, we have two fields, $\chi$ and $\Psi$, among which the first one does not propagate as there are no time-derivatives acting on it. By varying the action with respect to $\chi$, we find the following constraint:  
\begin{equation}\label{constraintPureR2}
    \Delta^2\chi=3\Delta\Box\Psi\qquad\to\qquad \Delta\chi=3\Box\Psi, 
\end{equation}
whose solution is given by: 
\begin{equation}\label{pureconstraint1}
    \chi=\frac{3}{\Delta}\Box \Psi. 
\end{equation}
Here, should think of the operator $\frac{1}{\Delta}$ in the sense of the Fourier transform. For the quantity 
\begin{equation}
    X(\Vec{x},t)=\int \frac{d^3k}{(2\pi)^{3/2}} X_{\Vec{k}}(t)e^{i\Vec{k}\Vec{x}},
\end{equation}
it is simply given by:
\begin{equation}
    \frac{1}{\Delta} X=\int \frac{d^3k}{(2\pi)^{3/2}} \frac{1}{|\Vec{k}|^2}X_{\Vec{k}}(t)e^{i\Vec{k}\Vec{x}}.
\end{equation}
By substituting the solution (\ref{pureconstraint1}) back to the Lagrangian density (\ref{LpureCPT3}), we find: 
\begin{equation}
    \mathcal{L}_{pure}=0. 
\end{equation}
In other words, the constraint (\ref{pureconstraint1}) has exactly cancelled the contribution of the field $\Psi$. This means that within this approach, inspired by the cosmological perturbation theory, we have found that the pure $R^2$ gravity has no propagating degrees of freedom at all. One should note that an equivalent result holds if we found the constraint for $\Phi$, solved it and substituted it back to the action. 
This result clearly contradicts the previous result, or even with the result that might follow from studying the theory in the Einstein or String frames. While in the appendix we show that the two transformation to the two frames is singular for flat space, before we discuss the reason for the discrepancy between the manifestly covariant and the direct approach, let us first study the remaining cases. 

\subsubsection{\textsc{The Full $R+R^2$ Gravity}}
Let us now generalize the previous procedure to the action that also contains an Einstein-Hilbert term: 
\begin{equation}
    S_{full}=\int d^4x\sqrt{-g}\left[\frac{M^2}{2}R+\beta R^2\right]. 
\end{equation}
Then, by perturbing the metric around the flat spacetime, we find the Lagrangian density given in (\ref{Lagrangianfull}). Decomposing the metric perturbations according to (\ref{decompositionCPT}), we find: 
\begin{equation}\label{fullCTPScalar}
    \mathcal{L}_{full}=\mathcal{L}_{full}^S+\mathcal{L}_{full}^V+\mathcal{L}_{full}^T,
\end{equation}
where
\begin{equation}\label{fullCTPVector}
    \mathcal{L}_{full}^S=M^2\left(2\Phi\Delta\Psi-3\dot{\Psi}\dot{\Psi}-\Psi\Delta\Psi\right)+4\beta\left(3\Ddot{\Psi}-2\Delta\Psi+\Delta\Phi\right)^2,
\end{equation}
\begin{equation}
   \mathcal{L}_{full}^V=-\frac{M^2}{4}V_i\Delta V_i,
\end{equation}
and 
\begin{equation}
    \mathcal{L}_{full}^T=-\frac{M^2}{8}\partial_{\alpha}h_{ij}^T\partial^{\alpha}h_{ij}^T.
\end{equation}
Here, the vector modes are expressed through an gauge-invariant variable: 
\begin{equation}
    V_i=S_i-\dot{F}_i. 
\end{equation}
We can notice that the theory now contains two massless tensor modes, two scalars and a vector. However, the vector modes are not propagating -- they do not have any kinetic terms. By varying the action with respect to $V_i$, we find the constraint: 
\begin{equation}
    \Delta V_i=0, 
\end{equation}
whose solution is 
\begin{equation}
    V_i=0. 
\end{equation}
Thus, we can set their correponsing Lagrangian density to zero. 

The Lagrangian density corresponding to the scalar modes is the most complicated one. However, similarly to the pure $R^2$ gravity, the gauge-invariant scalar potential $\Phi$ is not propagating. By varying the action with respect to it, we arrive at the following constraint: 
\begin{equation}
   4\beta\Delta\left(3\Ddot{\Psi}-2\Delta\Psi+\Delta\Phi\right)+M^2\Psi=0.
\end{equation}
Its solution is given by: 
\begin{equation}
    \Phi=\frac{1}{\Delta}\left(-3\Ddot{\Psi}+2\Delta\Psi-\frac{M^2}{\beta}\Psi\right). 
\end{equation}
By substituting it back to (\ref{fullCTPScalar}), we find: 
\begin{equation}
     \mathcal{L}_{full}^S=3M^2\left(\dot{\Psi}\Dot{\Psi}+\Psi\Delta\Psi-m_{\Psi}^2\Psi\Psi\right),
\end{equation}
where 
\begin{equation}\label{StBScalarMassCTP}
    m_{\Psi}^2=\frac{M^2}{12\beta}. 
\end{equation}

Thus, Starobinsky model in flat spacetime describes a massive scalar degree of freedom, with mass given by (\ref{StBScalarMassCTP}), and two massless tensor modes, in agreement with the manifestly covariant approach. 


\section{\Large\textsc{Generalization to the 3-form }}

In the previous sections, we have seen that the two approaches -- the manifestly covariant one, and the direct one -- yield different results for the pure $R^2$ gravity. The first method indicates that the theory has a single scalar \textit{dof}, while the second one results in no \textit{dof} at all. 

However -- \textit{How special is $R^2$ gravity?} As we will see in this section, there exists at least one more example where the two methods provide different results -- the theory of a massless 3-form. Its action is given by: 
\begin{equation}\label{3formaction}
    S=-\frac{1}{48}\int d^4x W_{\mu\nu\alpha\beta}W^{\mu\nu\alpha\beta},
\end{equation}
where 
\begin{equation}
    W_{\mu\nu\alpha\beta}=C_{\nu\alpha\beta,\mu}-C_{\mu\alpha\beta,\nu}+C_{\beta\mu\nu,\alpha}-C_{\alpha\mu\nu,\beta}
\end{equation}
is the corresponding field strength. It is well known that if the three form is massless, it describes no \textit{dof}, while if massive, it propagates one pseudoscalar \textit{dof} \cite{Kawai, Trugenberger, Quevedo, 2001Smailagic, 2002Casini, Auria, Buchbinder, Dalmazi, Shifman, Garcia, Kuzenko, CF1980, C2019, Dvali}. Notably, the \textit{dof} can be easily obtained using the direct approach as in \cite{Hell:2021wzm}. In order to demonstrate the discrepancy that occurs for this theory when studied via the two approaches, we will rewrite the theory in terms of the vector field, dual to the 3-form: 
\begin{equation}
    A_{\mu}=\varepsilon_{\mu\nu\rho\sigma}C^{\nu\rho\sigma}. 
\end{equation}
In terms of the vector field, (\ref{3formaction}) becomes: 
\begin{equation}\label{VfieldAction}
    S=\frac{1}{2}\int d^4x \left(\partial_{\mu}A^{\mu}\right)^2.
\end{equation}
Let us now study this action via the two different approaches. 

\subsection{\large\textsc{The Manifestly Covariant Approach}} 

In this subsection, we will focus on the manifestly covariant approach. For this, let us write the vector field \textit{a la Stueckelberg}: 
\begin{equation}\label{vectorStueckelberg}
    A_{\mu}=A_{\mu}^T+\partial_{\mu}\phi. 
\end{equation}
As in the case of $R^2$ gravity, this introduces a new field to the theory -- the scalar field $\phi$. This is clear, as the vector field is not divergenceless at this moment. However, by writing the original field as (\ref{vectorStueckelberg}) introduces also a gauge redundancy: 
\begin{equation}\label{gredvecfield}
A^T_{\mu}\to\Tilde{A}_{\mu}=A^T_{\mu}+\partial_{\mu}\lambda\qquad\text{and}\qquad \phi\to\Tilde{\phi}=\phi-\lambda.
\end{equation}
Let us use it to set the Lorentz gauge. For this, we can choose the gauge parameter $\lambda$, such that 
\begin{equation}\label{LorentzcondExapmple}
    \partial^{\mu}\Tilde{A}_{\mu}=0. 
\end{equation}
This is possible for 
\begin{equation}
    \Box\lambda=-\partial_{\mu}A^{T\mu},
\end{equation}
and fixes the gauge, up to an additional parameter $\gamma$:
\begin{equation}\label{resgaugeredvector}
A^T_{\mu}\to\Tilde{A}_{\mu}=A^T_{\mu}+\partial_{\mu}\gamma\qquad\text{and}\qquad \phi\to\Tilde{\phi}=\phi-\gamma.
\end{equation}
which satisfies 
\begin{equation}
    \Box\gamma =0. 
\end{equation}
Then, the above approach (\ref{vectorStueckelberg}) is precisely the vector analogue of the Stueckelberg trick for $R^2$ gravity that we have performed in the first section.  

In this case, the Lagrangian density corresponding to the action (\ref{VfieldAction}) becomes: 
\begin{equation}\label{LdenVecMCOV}
    \mathcal{L}=\frac{1}{2}\Box \phi\Box\phi, 
\end{equation}
and leads to the following equation for the scalar field: 
\begin{equation}
    \Box^2\phi=0. 
\end{equation}
Even thought this equation is fourth-order in time-derivatives, it nevertheless does not describe two degrees of freedom -- a ghost and a healthy one. Due to the residual gauge redundancy (\ref{resgaugeredvector}), one degree of freedom can be removed, so the resulting theory only describes only one degree of freedom. However, based on the literature to date, this is an incorrect result -- the theory described by the action (\ref{VfieldAction}) should not have any \textit{dof}. In order to convince ourselves that this is the case, let us study the theory directly. 

\subsection{\large\textsc{Using the Methods of the Cosmological Peturbation Theory}}
Let us now approach the action (\ref{VfieldAction}) by using the direct approach, that is based on the cosmological perturbation theory, as we have done for the $R^2$ gravity. We will separate the time and spatial component of the vector field $(A_0, A_i)$, and further decompose: 
\begin{equation}
    A_i=V_i^T+\partial_i\chi, \qquad \text{with}\qquad V_{i,i}^T=0.
\end{equation}
Then, we find:
\begin{equation}\label{LdenVecNCOV}
    \mathcal{L}=\frac{1}{2}\left(\dot{A_0}\dot{A_0}-2\dot{A_0}-2\dot{A_0}\Delta\chi+\Delta\chi\Delta\chi\right).
\end{equation}
We can notice that the field $V_i^T$ has dropped out from the action. Moreover, similarly to the case of the pure $R^2$ gravity, the scalar $\chi$ is not propagating. It satisfies a constraint: 
\begin{equation}\label{constrainttoymodl}
    \Delta\chi=\dot{A}_0,
\end{equation}
whose solution is given by 
\begin{equation}
    \chi=\frac{1}{\Delta}\dot{A_0}.
\end{equation}
By substituting it back to (\ref{LdenVecNCOV}), we find 
\begin{equation}
    \mathcal{L}=0. 
\end{equation}
Thus, with the direct decomposition, we have confirmed that the this theory has no \textit{dof}, in agreement with the previous literature. 

\subsection{\large\textsc{The Massive Case}}
For completeness, let us also consider the massive case, whose action is given by: 
\begin{equation}
    S=\frac{1}{2}\int d^4x \left[\left(\partial^{\mu}A_{\mu}\right)^2 + m^2A_{\mu}A^{\mu}\right].
\end{equation}
Interestingly, as we will see, two approaches coincide in this case, and give the same number of degrees of freedom. 

On the one hand, in the direct approach, the Lagrangian becomes 
\begin{equation}
    \mathcal{L}=\frac{1}{2}\left[\dot{A_0}\dot{A_0}-m^2A_0A_0-2\dot{A_0}\Delta\chi+\Delta\chi(\Delta-m^2)\chi\right].
\end{equation}
The $\chi$ satisfies now a different constraint: 
\begin{equation}\label{constraintexamplechi}
    \Delta(\Delta-m^2)\chi-\Delta\dot{A}_0=0,
\end{equation}
whose solution is given by: 
\begin{equation}
    \chi=-\frac{1}{-\Delta+m^2}\dot{A}_0. 
\end{equation}
Upon subsituting it to the Lagrangian density, we find: 
\begin{equation}
    \mathcal{L}=-\frac{1}{2}A_0(-\Box+m^2)\frac{m^2}{-\Delta+m^2}A_0,
\end{equation}
which describes one massive scalar degree of freedom. 

Approaching to the theory, on the other hand, with the Stueckelberg trick supplemented with Lorentz gauge, we find: 
\begin{equation}
    \mathcal{L}=\frac{1}{2}\Box\phi(\Box-m^2)\phi
\end{equation}
thus resulting in one degree of freedom as well, as one of the two that would arise from the above Lagrangian density can be removed with the residual gauge redundancy (\ref{resgaugeredvector}).

\section{\Large\textsc{The Origin of the Discrepancy}}
So far, we have studied the \textit{dof} of $R^2$ gravity and the dual of a 3-form theory via two approaches -- the direct approach based on the tools of the cosmological perturbation theory, and the manifestly covariant one. We have found that while the two agree in the case of a massive 3-form, and the full $R^2$ gravity in flat spacetime, they give conflicting results for the massless 3-form, and the pure $R^2$ gravity. In particular: 

\textcolor{YellowOrange}{$\diamond$}\;\;  \textsc{\textbf{$R^2$ Gravity}}

For the $R^2$ gravity, we have found that the direct approach gives
\begin{equation}
\begin{split}
     &\mathcal{L}_{pure}=0,\qquad\text{and}\\\\
     &\mathcal{L}_{full}=3M^2\left(\dot{\Psi}\Dot{\Psi}+\Psi\Delta\Psi-m_{\Psi}^2\Psi\Psi\right)-\frac{M^2}{8}\partial_{\alpha}h_{ij}^T\partial^{\alpha}h_{ij}^T.
\end{split}
\end{equation}
In contrast, by performing the Stueckelberg trick and by implementing the Lorentz-like gauges, we find: 
\begin{equation}
\begin{split}
     &\mathcal{L}_{pure}=\frac{9\beta}{16}\Box\sigma\Box\sigma,\qquad\text{and}\\\\
     &  \mathcal{L}_{full}=-\frac{M^2}{8}\partial_{\alpha}l_{\mu\nu}^T\partial^{\alpha}l^{T\mu\nu}+\frac{9\beta}{16}\Box\sigma\left(\Box-m_{\sigma}^2\right)\sigma.
\end{split}
\end{equation}

\textcolor{YellowOrange}{$\diamond$}\;\;  \textsc{\textbf{3-form Dual }}

The two approaches also do not agree in the case of a dual of a 3-form. There, we have found that the direct approach leads to: 

\begin{equation}
    \begin{split}
        &\mathcal{L}=0,\qquad\text{for}\qquad m=0,\qquad \text{and}\\\\
        &\mathcal{L}=-\frac{1}{2}A_0(-\Box+m^2)\frac{m^2}{-\Delta+m^2}A_0,\qquad\text{for}\qquad m\neq 0.
    \end{split}
\end{equation}
Applying the Stueckelberg trick with the Lorentz gauge, we find: 
\begin{equation}
    \begin{split}
        &\mathcal{L}=\frac{1}{2}\Box \phi\Box\phi,\qquad\text{for}\qquad m=0,\qquad \text{and}\\\\
        &\mathcal{L}=\frac{1}{2}\Box\phi(\Box-m^2)\phi,\qquad\text{for}\qquad m\neq 0.
    \end{split}
\end{equation}

Let us now compare the two approaches and discuss what might be the cause of this discrepancy. The manifestly covariant approach is appealing as the manifest Lorentz covariance is preserved. It relies on the Stueckelberg trick, in which one has to introduce additional fields. In the case of a vector field, dual to the 3-form, we have introduced an additional scalar, while in the gravity case, we introduced both scalar and vector fields. However, this comes at a cost -- the additional fields are introduced such that the system has additional gauge redundancy. Thus, even though the scalar fields appear with higher-time derivatives, the system has no ghost degrees of freedom -- they can be removed using the residual redundancy. 

From this perspective, the direct approach, which relies on the tools of the cosmological perturbation theory could be more appealing. While the resulting Lagrangian is not manifestly covariant, the apparent ghosts do not appear and the theory has and no gauge redundancy. It also provides a straightforward way to quantize the (linearized) theory. One simply promotes the fields $\Psi$ and $h_{ij}^T$, and their corresponding moments
\begin{equation}
    \pi_{\Psi}=6M^2\dot\Psi\qquad\text{and}\qquad \pi^T_{ij}=\frac{M^2}{4}\dot h_{ij}^T
\end{equation}
to operators, and postulates the corresponding canonical commutation relations. 

However, the two approaches do not give the same number of \textit{dof} for the pure $R^2$ gravity or the massless 3-form. In the manifestly covariant approach, one could always pick the unitary gauge, which sets the additionally introduced fields to zero. 
Therefore, we can conclude that the problem might not arise by just performing the Stueckelberg trick -- we can always go back to the gauge-invariant variables and turn to the direct approach. 

One can wonder if the problem of counting the \textit{dof} lies in the direct approach, due to the substitution of the equations (\ref{constraintPureR2}) and (\ref{constrainttoymodl}) back to the action. However, these equations are constraints, and will thus be valid at all times. This can be easily seen from the Hamiltonian formalism, that we have presented in the appendix. There, we have found that it yields the same number of \textit{dof} as we have found with the direct approach.  

This leads us to the conclusion that the reason for this discrepancy lies in the Lorentz-like gauge conditions, given in (\ref{MCOVproperties1}) and (\ref{LorentzcondExapmple}). By imposing them, it seems that there is not enough gauge redundancy to remove all degrees of freedom and provide matching between the two approaches. 

This is similar to the problem with the synchronous gauge, arising in the cosmological perturbation theory.  Following \cite{Cosmo}, let us consider the scalar perturbations of the background metric for the Friedmann-Robertson-Lemetre-Walker Universe. In this case, the metric given by:
\begin{equation}
    ds^2=a^2(\eta)\left[\left(-1+2\phi\right)d\eta^2+2\partial_i B dx^id\eta+\left((1+2\psi)\delta_{ij}+2\partial_i\partial_j E\right)dx^idx^j\right].
\end{equation}
Under infinitesimal coordinate transformations: 
\begin{equation}
    x^{\mu}\to\Tilde{x}^{\mu}=x^{\mu}+\xi^{\mu},
\end{equation}
the above potentials transform as 
\begin{equation}
   \begin{split}
        &\phi\to\Tilde{\phi}=\phi-\frac{a'}{a}\xi_0-\xi_0'\qquad\qquad B\to\Tilde{B}=B-\zeta'-\xi_0\\
        &\psi\to\Tilde{\psi}=\psi+\frac{a'}{a}\xi_0\qquad\quad\text{and}\qquad\qquad E\to\Tilde{E}=E-\zeta. 
   \end{split}
\end{equation}
Let us now first consider the conformal gauge: 
\begin{equation}
    \Tilde{E}=0\qquad\text{and}\qquad \Tilde{B}=0. 
\end{equation}
It is easy to show that this choice fixes the coordinate system uniquely. If $E$ and $B$ are zero in one coordinate frame, the gauge parameters are exactly determined: 
\begin{equation}
\zeta=0\qquad \text{and}\qquad \xi^0=0.
\end{equation}
The same is not true, however, in the synchronous gauge, defined by: 
\begin{equation}\label{synchronousgauge}
    \Tilde{\phi}=0\qquad\text{and}\qquad \Tilde{B}=0. 
\end{equation}
It is possible to make a gauge transformation to this frame if
\begin{equation}
    \phi=\frac{1}{a}(a\xi_0)'\qquad \qquad B=\xi_0+\zeta'. 
\end{equation}
However, if $\phi$ and $B$ are then zero in one frame of reference, there is a whole class of frames where this will also be true, all connected with that one by: 
\begin{equation}
    \Tilde{\eta}=\eta+\frac{C_1(\Vec{x})}{a}\qquad\text{and}\qquad \Tilde{x}^i=x^i+\partial^i C_2(\Vec{x})+\int d\eta\frac{\partial^i C_1 (\Vec{x})}{a}. 
\end{equation}
In other words, (\ref{synchronousgauge}) does not fix the degrees of freedom uniquely -- the constants of integration prevent this. 
As a result, the synchronous gauge can give rise to fictitious, unphysical modes. 

Let us now draw a parallel with our previous example. In the case of the vector field, the Stueckelberg decomposition introduced a gauge redundancy:
\begin{equation}
    A_{\mu}^T\to\Tilde{A}_{\mu}^T =A_{\mu}^T+\partial_{\mu}\lambda\qquad\text{and}\qquad \phi\to\Tilde{\phi}=\phi-\lambda. 
\end{equation}
It is easy to see that the Coulomb gauge:
\begin{equation}
    \partial_i\Tilde{A}_i=0
\end{equation}
fixes the gauge uniquely. In this case
\begin{equation}
    \Delta\lambda=-\partial_iA_i=-\Delta\chi\qquad\to\qquad\lambda=-\chi.
\end{equation}
Here we have decomposed $A_i=A_i^T+\partial_i\chi$. If now $A_{\mu}^T$ and $\phi$ are zero, it implies that $\lambda=0$, and thus our gauge choice uniquely fixes the gauge. A similar thing happens with the unitary gauge, in which 
\begin{equation}
    \Tilde{\phi}=0. 
\end{equation}
However, unlike the previous two cases, the Lorentz gauge does not fix the gauge uniquely: 
\begin{equation}
    \partial_{\mu}\Tilde{A}^{\mu}=0\qquad\to \qquad \Box\lambda=-\partial_{\mu}A^{\mu}. 
\end{equation}
If we set in one frame now $A_{\mu}=0$, it implies
\begin{equation}
    \Box\lambda=0
\end{equation}
and thus 
\begin{equation}
    \lambda_k=C_1e^{ikt}+C_2e^{-ikt} 
\end{equation}
in the Fourier space. The appearance of the constants of integration shows that the Lorentz gauge is not uniquely fixed. Following the analogy with the synchronous gauge, the result of this is the presence of fictitious degrees of freedom. 

This can also be seen from the fact that the gauge invariant fields (obtained in the direct approach) are connected with those of the Stueckelberg trick by 
\begin{equation}
    \Delta\chi-\dot A_0=\partial_{\mu}A^{T\mu}+\Box\phi
\end{equation}
By applying the constraint (\ref{constrainttoymodl}) as well as the Lorentz gauge, we find:
\begin{equation}
    0=\Box\phi. 
\end{equation}
Thus, where gauge invariant variables give us nothing -- the scalar gives us something -- it is the fictitious mode, an artefact of the gauge choice. The same analysis can also be applied to the $R^2$ gravity, by a complete analogy. Thus, by performing the manifestly covariant procedure, which consists of the Stueckelberg trick along with the Lorentz-like conditions to partially fix the gauge, one can find fictitious modes in a similar way as in the synchronous gauge of the cosmological perturbation theory. 

\section{\Large\textsc{Beyond the Linear Approximation}}

So far, we have seen that the full $R^2$ gravity propagates a massive scalar mode and a massless tensor mode. Yet, by using a gauge invariant approach, we have seen that in the pure theory, these modes are absent. This brings us to the following question -- \textit{ Is there a discontinuity between the two theories when each of the parameters of the model vanishes? } 

In this section, we will study the full $R^2$ theory, described by the action (\ref{StarobinskyAction}), and show that the discrepancy in the number of \textit{dof}is just an artefact of the perturbation theory. In particular, we will investigate the limit when $M\to0$,  by following the procedure along the lines of \cite{Chamseddine:2018gqh, Hell:2021oea, Hell:2021wzm} which searches for \textit{the strong coupling scale}. The main ingredient for this analysis is the minimal amplitude of the quantum fluctuations of the fields -- a direct consequence of Heisenberg's uncertainty principle.

\subsection{\textsc{The Free Theory and the Quantum Fluctuations}}

In this subsection, we will find the amplitude of the quantum fluctuations of the propagating modes of the full $R^2$ theory, studying each of them separately. 

\subsubsection{\textsc{\textbf{Scalar Modes}}} 

By expanding (\ref{StarobinskyAction}) in terms of the metric perturbations, solving the constraints, and substituting them back to the action, we have previously found that the Lagrangian density describing the scalar modes of the linear theory is given by: 
\begin{equation}
    \mathcal{L}_S=6M^2\left(\dot\psi\dot\psi+\psi\Delta\psi-m_{\psi}^2\psi^2\right),
\end{equation}
where 
\begin{equation}
    m_{\psi}^2=\frac{M^2}{6\beta}
\end{equation}
is the mass of the scalar field. As we have previously pointed out, we can notice that by setting $M=0$, this scalar mode drops out, giving the agreement with the pure $R^2$ gravity. 

In order to estimate the scale at which the above expansion will no longer hold, let us find the fluctuations of the perturbations. For the normalized scalar field, $\psi_n$, the minimal amplitude of quantum fluctuations on the scales $k\sim\frac{1}{L}$ is given by \cite{QFTCS}
\begin{equation}
    \delta\psi_n\sim\left.\sqrt{\frac{k^3}{\omega_k}}\right|_{k\sim\frac{1}{L}}
\end{equation}
In our case, the normalized scalar field is given by: 
\begin{equation}
    \psi_n=\sqrt{12} M \psi,
\end{equation}
while
\begin{equation}
    \omega_{\psi}^2=k^2+m_{\psi}^2. 
\end{equation}

Thus for energies $k^2\sim\frac{1}{L^2}\gg m_{\psi}^2$, the minimal amplitude of quantum fluctuations for the original scalar field is given by: 
\begin{equation}\label{psiqfluctA}
    \delta\psi\sim\frac{1}{ML}. 
\end{equation}

We should note that the limit $\beta\to0$ seems singular -- in that case, the mass of the scalar field diverges. Nevertheless, the above expression was obtained by expanding the metric around the Minkowski space-time: 
\begin{equation}\label{expansion}
    g_{\mu\nu}=\eta_{\mu\nu}+h_{\mu\nu},
\end{equation}
and holds\textbf{ only} for the metric fluctuations that are smaller than unity. Thus, at low energies $k^2\sim\frac{1}{L^2}\ll m_{\psi}^2$,  the minimal amplitude of quantum fluctuations for the scalar field:
\begin{equation}\label{psiqfluctB}
    \delta\psi\sim\frac{1}{ML\sqrt{L m_{\psi}}}\sim\frac{\beta^{1/4}}{\left(ML\right)^{3/2}}. 
\end{equation}
becomes of order of unity on length scales
\begin{equation}
    L_{\psi}\sim\frac{1}{(M\sqrt{m_{\psi}})^{2/3}}\sim \frac{\beta^{1/6}}{M}
\end{equation}
beyond which the above expansion (\ref{expansion}) no longer holds. 
Similarly, the expansion will not hold at high-energies, once the length scales
\begin{equation}
    L_{str}\sim\frac{1}{M},
\end{equation}
are reached.  

In the remaining part of this work, we will be interested in the high-energy limit -- the limit when the $R^2$ contribution dominates over the Einstein term.

\subsubsection{\textsc{\textbf{Tensor Modes}}} 

As we have previously found, the tensor modes are described by the following Lagrangian density:
\begin{equation}
    \mathcal{L}_T=-\frac{M^2}{4}\partial_{\alpha}h_{ij}^T\partial^{\alpha}h_{ij}^T. 
\end{equation}
We can see that similarly to the scalar modes, they vanish for $M=0$. The corresponding amplitude of quantum fluctuations for the tensor modes is given by 
\begin{equation}\label{tensorqfluct}
    \delta h^T_{ij}\sim\frac{1}{ML}. 
\end{equation}
Thus, once $ L\sim\frac{1}{M}$, the tensor modes enter a strong coupling regime and the metric expansion no longer holds (\ref{expansion}). 

\subsection{\large\textsc{The First-Order Corrections}}

Let us now consider the first-order corrections to the full $R^2$ gravity. By expanding the metric according to (\ref{expansion}), we find the following Lagrangian density involving the cubic terms of the metric perturbations: 
\begin{equation}
    \begin{split}
        \mathcal{L}=\mathcal{L}_0+\mathcal{L}_{int},
    \end{split}
\end{equation}
where 
\begin{equation}
    \mathcal{L}_0=\frac{M^2}{2}\left(2h^{\mu\nu}_{,\mu}h_{\alpha\nu}^{,\alpha}-h_{\mu\nu,\alpha}h^{\mu\nu,\alpha}-2h^{\mu\nu}_{,\mu}h_{,\nu}+h_{,\mu}h^{,\mu}\right)+\beta\left(h^{\mu\nu}_{,\mu\nu}-h_{,\mu}^{,\mu}\right)^2,
\end{equation}
and 
\begin{equation}
    \begin{split}
        \mathcal{L}_{int}&=\beta\left(h^{\alpha\beta}_{,\alpha\beta}-h_{,\alpha}^{,\alpha}\right)\left[2h^{\mu\nu}\left(h_{,\mu\nu}-2h^{\gamma}_{\nu,\mu\gamma}+h_{\mu\nu,\gamma}^{,\gamma}\right)+\frac{1}{2}h\left(h^{\mu\nu}_{,\mu\nu}-h_{,\mu}^{,\mu}\right)\right.\\&\left. 
        -\frac{1}{2}h_{,\mu}h^{,\mu}-2h_{,\mu}^{\mu\nu}h^{\gamma}_{\nu,\gamma}+2h^{,\mu}h^{\nu}_{\mu,\nu}-h_{\gamma\nu,\mu}h^{\gamma\mu,\nu}+\frac{3}{2}h_{\mu\nu,\gamma}h^{\mu\nu,\gamma}\right]\\
        &+M^2\left(-\frac{1}{4}h^{\alpha\beta}h_{\mu\nu}h^{\mu\nu}_{,\alpha\beta}-\frac{1}{8}h_{\mu\nu}h^{\mu\nu}h^{\alpha\beta}_{,\alpha\beta}-\frac{1}{4}h^{\alpha\beta}h_{\beta\nu}h_{\alpha,\delta}^{\nu,\delta}-h^{\alpha\beta}h_{\alpha,\mu}^{\mu}h^{\nu}_{\beta,\nu}\right.\\&\left. 
        +\frac{1}{2}h^{\alpha\beta}h_{\beta\nu,\mu}h^{\mu,\nu}_{\alpha}-h^{\alpha\beta}_{,\mu}h_{\alpha}^{\mu}h^{\nu}_{\beta,\nu}+\frac{1}{8}hh_{\mu\nu}h^{\mu\nu,\gamma}_{,\gamma}+\frac{3}{16}h_{\mu\nu}h^{\mu\nu}h_{,\gamma}^{,\gamma}+h^{\alpha\beta}h_{\alpha,\mu}^{\mu}h_{,\beta}\right.\\&\left.+\frac{1}{2}hh^{\mu\nu}_{,\mu}h^{\alpha}_{\nu,\alpha}-\frac{1}{4}hh_{\alpha\nu,\mu}h^{\alpha\mu,\nu}+\frac{1}{4}hh^{\alpha\beta}h_{,\alpha\beta}-\frac{1}{16}h^2h^{,\mu}_{,\mu}\right)
    \end{split}
\end{equation}

In the above relations, comma denotes the derivative: $_{,\mu}=\partial_{\mu}$. For example: 
\begin{equation*}
    h_{,\mu}=\partial_{\mu}h\qquad\text{and}\qquad h_{,\mu\nu}=\partial_{\mu}\partial_{\nu}h,
\end{equation*}
where $h=h^{\mu}_{\mu}$. Let us now decompose the metric perturbations according to (\ref{decompositionCPT}). For simplicity, we will impose the conformal gauge: 
\begin{equation}
 E=0\qquad \text{and}\qquad B=0,   
\end{equation}
and use the residual gauge redundancy to set
\begin{equation}
    F_i=0. 
\end{equation}
As in the free case, the action corresponding to the above Lagrangian density has two constraints -- one for the scalar mode $\phi$, and one for the vector modes $S_i$, as these fields are not propagating. These constraints can be found by varying the action with respect to the two modes. By finding them, solving them perturbatively, and substituting them back to the action, we find the following Lagrangian density up to quartic terms in metric perturbations: 
\begin{equation}
    \begin{split}
        \mathcal{L}=\mathcal{L}_0+\mathcal{L}_{int},
    \end{split}
\end{equation}
where
\begin{equation}
    \mathcal{L}_0=6M^2\left(\dot\psi\dot\psi+\psi\Delta\psi-m_{\psi}^2\psi^2\right)-\frac{M^2}{4}h_{ij,\alpha}^Th^{T,\alpha}_{ij},
\end{equation}
and 
\begin{equation}\label{origigiLint}
    \begin{split}
        \mathcal{L}_{int}=&M^2\left\{-3\Ddot{\psi}\psi^2-15\psi^2\Delta\psi+12m_{\psi}^2\psi^3-\frac{1}{4}\psi\dot{h}_{ij}^T\dot{h}_{ij}^T+\frac{1}{2}h_{ij}^Th_{ij}^T\Ddot{\psi} -4\psi h_{ij}^T\Delta h_{ij}^T\right.\\&\left.
       -\frac{9}{4}\psi h_{ij,k}^Th_{ij,k}^T+\frac{3}{2}\psi h_{ij,k}^Th_{ik,j}^T-\frac{1}{4}h_{ij}^Th_{ik}^Th_{jk,\mu}^{T,\mu}-\frac{1}{4}h_{ij}^Th_{kl}^Th_{kl,ij}^T+\frac{1}{2}h_{ij}^Th_{il,k}^Th_{jk,l}^T\right.\\&\left.
       -6\left[3\dot\psi\dot\psi-2\psi\Delta\psi+\psi_{,i}\psi_{,i}+3m_{\psi}^2\psi^2+h_{ij}^T\psi_{,ij}+\frac{1}{8}\dot{h}_{ij}^T\dot{h}_{ij}^T-\frac{1}{2}h_{ij}^T\Delta h_{ij}^T\right.\right.\\&\left.\left.-\frac{3}{8}h_{ij,k}^Th_{ij,k}^T+\frac{1}{4}h_{ij,k}^Th_{ik,j}^T\right]\frac{1}{\Delta}\left[\Ddot{\psi}+m_{\psi}^2\psi\right]\right\}
    \end{split}
\end{equation}

We can see that by setting $M=0$, all of the terms of the Lagrangian density disappear at the cubic order\footnote{One can easily confirm that this holds even if one would not set the conformal gauge, and $F_i=0$.}. This gives further support for $R^2$ having no dof, now confirmed also at the cubic order. 

\subsection{\large\textsc{The Strong Coupling}}

In order to find the strong coupling scale, let us find the equations of motion for the scalar and tensor modes. By varying the action with respect to $\psi$ and $h_{ij}^T$, and expanding the two as
\begin{equation}
    \psi=\psi^{(0)}+\psi^{(1)}+...\qquad\text{and}\qquad h_{ij}^T=h_{ij}^{T(0)}+h_{ij}^{T(1)}+..
\end{equation}
where the zeroth values satisfy:
\begin{equation}
    (-\Box+m_{\psi}^2)\psi^{(0)}=0\qquad\text{and}\qquad \Box h_{ij}^{T(0)}=0,
\end{equation}
we find that the first-order contributions to the scalar and tensor modes satisfy respectively: 
\begin{equation}\label{eompsi1}
   \begin{split}
        12(-\Box+m_{\psi}^2)\psi^{(1)}&=22\dot{\psi}^{(0)}\dot{\psi}^{(0)}-24m_{\psi}^2\psi^{(0)}\psi^{(0)}+12\psi^{(0)}\Delta\psi^{(0)}+6\psi^{(0)}_{,i}\psi^{(0)}_{,i}
        \\&-6h_{ij}^{T(0)}\psi^{(0)}_{,ij}+\frac{3}{4}\dot{h}_{ij}^{T(0)}\dot{h}_{ij}^{T(0)}-3h_{ij}^{T(0)}\Delta h_{ij}^{T(0)}-\frac{9}{4}h_{ij,k}^{T(0)}h_{ij,k}^{T(0)} \\&+\frac{3}{4}h_{ij,k}^{T(0)}h_{ik,j}^{T(0)}
        -6\frac{1}{\Delta}\left[4\Delta\psi^{(0)}\Delta\psi^{(0)}-2\psi^{(0)}\Delta^2\psi^{(0)}+2\psi^{(0)}_{,i}\Delta\psi^{(0)}_{,i}\right.\\
        &\left.-4m_{\psi}^2\psi^{(0)}\Delta\psi^{(0)}-m_{\psi}^2\psi^{(0)}_{,i}\psi^{(0)}_{,i}+3m_{\psi}^4\psi^{(0)}\psi^{(0)}+3m_{\psi}^2\dot{\psi}^{(0)}\dot{\psi}^{(0)} \right.\\
        &\left.+\dot{\psi}^{(0)}_{,ij}\dot{h}_{ij}^{T(0)}+\psi^{(0)}_{,ij}\Delta h_{ij}^{T(0)} +h_{ij}^{T(0)}\Delta \psi^{(0)}_{,ij}-\frac{5}{8}\dot{h}_{ij}^{T(0)}\dot{h}_{ij}^{T(0)}\right.\\ &\left.-\frac{1}{4} \Delta h_{ij}^{T(0)}\Delta  h_{ij}^{T(0)}-\frac{1}{2} h_{ij}^{T(0)}\Delta^2 h_{ij}^{T(0)} -\frac{1}{4} h_{ij,k}^{T(0)}\Delta  h_{ij,k}^{T(0)}
        \right.\\&\left.+m_{\psi}^2\left(\frac{1}{8}\dot{h}_{ij}^{T(0)}\dot{h}_{ij}^{T(0)}-\frac{1}{2}h_{ij}^{T(0)}\Delta h_{ij}^{T(0)}-\frac{3}{8}h_{ij,k}^{T(0)}h_{ij,k}^{T(0)}+\frac{1}{4}h_{ij,k}^{T(0)}h_{ik,j}^{T(0)}\right)
        \right],
   \end{split}
\end{equation}
and
\begin{equation}\label{eomT1}
    \begin{split}
        -\Box h_{mn}^{(1)}&=2P_{mnij}^T\left(-m_{\psi}^2\psi^{(0)}h_{ij}^{T(0)}+2\dot{\psi}^{(0)}\dot{h}_{ij}^T-2\psi_{,k}^{(0)}h_{ij,k}^{T(0)}-6\psi^{(0)}\psi_{,ij}^{(0)}\right.\\&\left.
        +\frac{1}{2}\dot{h}_{il}^{T(0)}\dot{h}_{jl}^{T(0)}-\frac{1}{4}h_{kl}^{T(0)}h_{kl,ij}^{T(0)}+h_{il,k}^{T(0)}h_{jk,l}^{T(0)}-\frac{1}{2}h_{kl}^{T(0)}h_{ij,kl}^{T(0)}-h_{il,k}^{T(0)}h_{lk,j}^{T(0)}\right),
    \end{split}
\end{equation}
where 
\begin{equation}
   \begin{split}
        P_{mnij}^T&=\frac{1}{2}\left(\delta_{im}\delta_{jn}+\delta_{jm}\delta_{in}\right)-\frac{1}{2}\delta_{ij}\left(\delta_{mn}-\frac{\partial_k\partial_l}{\Delta}\delta_{km}\delta_{ln}\right)\\
        &+\frac{1}{\Delta}\left[\frac{1}{2}\delta_{mn}\partial_i\partial_j+\frac{1}{2}\delta_{mk}\delta_{ln}\frac{\partial_i\partial_j\partial_k\partial_l}{\Delta}-\delta_{im}\delta_{ln}\partial_l\partial_j-\delta_{jm}\delta_{ln}\partial_l\partial_i\right]
   \end{split}
\end{equation}
is the transverse-traceless projector.

Let us now first consider the scalar modes. At high energies, $k^2\sim\frac{1}{L^2}\gg m_{\psi}^2$, we can represent the most important terms for the equation of motion of the scalar modes (\ref{eompsi1}) as: 
\begin{equation}
   \begin{split}
        &12(-\Box+m_{\psi}^2)\psi^{(1)}\sim \frac{1}{L^2}\psi^{(0)}\psi^{(0)}+\frac{1}{L^2}\psi^{(0)}h_{ij}^{T(0)}+\frac{1}{L^2}h_{ij}^{T(0)}h_{ij}^{T(0)} 
   \end{split}
\end{equation}
where we have evaluated the derivatives as
\begin{equation}
    \partial_{\mu}\sim\frac{1}{L}. 
\end{equation}
By taking into account the minimal level of quantum fluctuations for the scalar and tensor modes (\ref{psiqfluctA}) and (\ref{tensorqfluct}), we can in addition estimate: 
\begin{equation}
    \psi^{(0)}\sim\frac{1}{ML} \qquad \text{and}\qquad h_{ij}^{T(0)}\sim\frac{1}{ML}. 
\end{equation}
Then, we find: 
\begin{equation}
    \psi^{(1)}\sim\frac{1}{\left(ML\right)^2}
\end{equation}
Clearly, in the limit $M\to0$, this term is divergent. However, once it becomes of the same order as the linear term, $\psi^{(0)}$, the scalar mode becomes strongly coupled, and thus we can no longer use the perturbative expansion. This corresponds to the length scales: 
\begin{equation}
    L_{str}\sim\frac{1}{M}. 
\end{equation}

Similarly, in the case of tensor modes, all the terms apart from the first one on the r.h.s. of (\ref{eomT1}) are equally dominant. Thus, we can estimate: 
\begin{equation}
    h_{ij}^{T(1)}\sim\frac{1}{\left(ML\right)^2}. 
\end{equation}
By comparing this non-linear term with the linear one, $h_{ij}^{T(0)}$, we find that the tensor modes enter the strong coupling regime at length scales
\begin{equation}
    L_{str}\sim\frac{1}{M}, 
\end{equation}
same as the scalar modes. 

Thus, we have found that on energies larger than the mass of the scalar field, the perturbative expansion holds for the length-scales $L>L_{str}$.\footnote{One can infer this by searching for which scales the first-order contributions are subdominant when compared to the linear terms.} However, once the strong coupling scale $L_{str}$ is reached, both scalar and tensor modes become strongly coupled. For completeness, in the Appendix we comment on the limit when $\beta\to0$.

\section{\Large\textsc{Discussion}}

The main purpose of this paper was to study the \textit{dof} of $R^2$ gravity in flat space-time and obtain an action that is given just in terms of the gauge-invariant variables. By performing an analysis based on the cosmological perturbation theory, we have found a surprising result -- the pure $R^2$ gravity has no \textit{dof}, in contradiction to the previous statements in the literature. We have also confirmed that the existing manifestly covariant procedure, based on the Stueckelberg trick and Lorentz-like conditions, yields the presence of one massless scalar in the theory. Moreover, we have shown that an analogous analysis leads to the same conflict for the vector field dual to the 3-form. 

However, as we have seen, similarly to the synchronous gauge, the manifestly covariant approach does not fix the gauge uniquely. As a result, the apparent scalar mode is fictitious,  leading us to conclude that the massless 3-form and pure $R^2$ gravity
have no \textit{dof}. This result is supported by the Hamiltonian formalism as well, which we have presented in the appendix. In addition, we have shown that the first-order non-linear terms disappear for $M=0$, supporting this result as well. 

Curiously, the scalar sector of $R^2$ gravity and the 3-form theory bear striking resemblance. If the Einstein-Hilbert term is added to the $R^2$ term, the scalar mode becomes massive. In its absence, however, the theory has no \textit{dof}. Similarly, if the 3-form is massive, it contains one massive pseudoscalar, while if massless no \textit{dof} remains. It would be interesting to investigate if such similarity between the two theories would persist once one takes into account the coupling between the 3-form and GR to include the tensor modes, or when the non-linear terms and the external matter are taken into account. 

The difference between the \textit{dof} of the pure and full $R^2$ gravity implies that there could be a discontinuity between the two theories. This would in turn imply that the perturbative series becomes singular when $M\to0$. However, the analysis of the first-order corrections suggests otherwise, implying that the discontinuity is just an artefact of the perturbation theory. 
In particular, we have found that both scalar and tensor modes are strongly coupled for length-scales $L\leq L_{str}\sim\frac{1}{M}$. This is also intuitive -- the Einstein term dominates over the $R^2$ contribution. One should note that this analysis also supports the absence of the degrees of freedom in beyond the linear approximation. If $M=0$, all terms in the Lagrangian vanish. It will be interesting to further study this theory in the presence of an external source, and investigate further similarities with the Vainstein mechanism.

\begin{center}
    \large\textsc{\textbf{Acknowledgements}}

\end{center}

\textit{The authors would like to thank Slava Mukhanov and Luis Álvarez-Gaumé for very useful discussions, Alex Kehagias and Costas Bachas for useful correspondence. In addition, A.H. would like to thank Elisa G. M. Ferreira and Misao Sasaki for motivating discussions and very useful suggestions, and Jun'ichi Yokoyama useful questions.  G.Z. would also like to thank MPP-Munich, CERN-TH and ITP-Heidelberg for
hospitality, and MPP-Munich, CERN-TH and DFG Exzellenzcluster
2181:STRUCTURES of Heidelberg University for support. The work of A. H. was supported by World Premier International Research Center Initiative (WPI), MEXT, Japan.
 The work of D.L.~is supported by the Origins Excellence Cluster and by the German-Israel-Project (DIP) on Holography and the Swampland.
}

\section*{\Large\textsc{Appendix }}
\addcontentsline{toc}{section}{\protect\numberline{}\Large\textsc{Appendix }}
\subsection*{\large\textsc{\textcolor{YellowOrange}{A} }\;\; \textsc{Hamiltonian Formalism for the Massless 3-form}}
\addcontentsline{toc}{subsection}{\protect\numberline{}\large\textsc{\textcolor{YellowOrange}{A} }\;\; \textsc{Hamiltonian Formalism for the Massless 3-form}}
In this Appendix we will study the Hamiltonian formalism for the massless 3-form and the pure $R^2$ gravity by following closely the constraint procedure described in \cite{Dirac}.  In this subsection, let us first consider the simplest case of the dual of the massless 3-form. The starting point is the Lagrangian density: 
\begin{equation}
    \mathcal{L}=\frac{1}{2} \left(\partial_{\mu}A^{\mu}\right)^2.
\end{equation}
The conjugated momenta are given by: 
\begin{equation}
    \pi^0=\dot A_0 -\Delta\chi
\end{equation}
and 
\begin{equation}
    \pi_{\chi}=0. 
\end{equation}
Since the momenta corresponding to the scalar $\chi$ is vanishing, it implies the existence of a primary constraint: 
\begin{equation}\label{TMC1}
    \pi_{\chi} \approx 0. 
\end{equation}
Then, the total Hamiltonian density is given by: 
\begin{equation}
    \mathcal{H}_T=\frac{1}{2}\pi^0\pi^0+\pi^0\Delta\chi+v\pi_{\chi},
\end{equation}
where $v$ is the Lagrange multiplier. 
Let us define the equal-time Poisson brackets: 
\begin{equation}
    \left\{f(\Vec{x},t),g(\Vec{y},t)\right\}=\sum_{i=1,2}\int d^3z\left( \frac{\delta f}{\delta Q_i(z)}\frac{\delta g}{\delta P_i(z)}-\frac{\delta g}{\delta Q_i(z)}\frac{\delta f}{\delta P_i(z)}\right),
\end{equation}
where,
\begin{equation}
    \begin{split}
        Q_1=A_0,\qquad P_1=\pi^0,\qquad Q_2=\chi,\qquad\text{and}\qquad P_2=\pi_{\chi}. 
    \end{split}
\end{equation}
Then, the equations of motion for the function $f$ are given by: 
\begin{equation}
    \dot f= \left\{f,H_T\right\},
\end{equation}
and we have: 
\begin{equation}
    \left\{A_0(\Vec{x},t),\pi^0(\Vec{y},t)\right\}=\delta^{(3)}(\Vec{x}-\Vec{y})\qquad\text{and}\qquad  \left\{\chi(\Vec{x},t),\pi_{\chi}(\Vec{y},t)\right\}=\delta^{(3)}(\Vec{x}-\Vec{y}). 
\end{equation}
From the consistency condition: 
\begin{equation}
    \dot \pi_{\chi}=0,
\end{equation}
we find the secondary constraint: 
\begin{equation}\label{TMC2}
    C=-\Delta \pi^0\approx 0\qquad\to\qquad \pi^0\approx 0.
\end{equation}
Since $\dot C=0$ implies $0=0$, we have thus found all of the constraints of the system. It is easy to see that the constraints (\ref{TMC1}) and (\ref{TMC2}) are first-class constraints -- their Poisson brackets vanish. 
The extended Hamiltonian -- quantity from which we can find the most general equations of motion of the above theory -- is then given by: 
\begin{equation}
    H_E=\int d^3x\left[\frac{1}{2}\pi^0\pi^0+\pi^0\Delta\chi+v\pi_{\chi}+u\pi^0\right]. 
\end{equation}
Let us now count the \textit{dof}. The equations of motion 
\begin{equation}
     \begin{split}
          &\dot \pi^0= \left\{\pi^0,H_E\right\}\qquad  \dot A_0= \left\{A_0,H_E\right\}\\
          &\dot \pi_{\chi}= \left\{\pi_{\chi},H_E\right\}\qquad  \dot \chi= \left\{\chi,H_E\right\}
     \end{split}
\end{equation}
require four initial conditions, and thus give us two degrees of freedom. Substracting the two constraints 
\begin{equation}
    \pi_{\chi}=0\qquad\text{and}\qquad \pi^0=0,
\end{equation}
cancels the two, thus thus we ultimately find that this theory has no propagating \textit{dof}, in agreement with the direct approach.

\subsection*{\large\textsc{\textcolor{YellowOrange}{B} }\;\; \textsc{Hamiltonian Formalism for the Pure $R^2$ Gravity}}
\addcontentsline{toc}{subsection}{\protect\numberline{}\large\textsc{\textcolor{YellowOrange}{B} }\;\; \textsc{Hamiltonian Formalism for the Pure $R^2$ Gravity}}
Let us now generalize the previous procedure to the case of the (linearized) pure $R^2$ gravity. Our starting point will be the Lagrangian density: 
\begin{equation}
    \begin{split}
\mathcal{L}_{pure}&=4\beta\left[\Delta\chi\Delta\chi-2\Delta\chi\Box\Psi_n+\Box\Psi_n\Box\Psi_n\right],
    \end{split}
\end{equation}
were we have rescaled 
\begin{equation}
    \Psi_n=3\Psi
\end{equation}
from (\ref{LpureCPT3}) for simplicity. Following the method described in \cite{Woodard:2015zca, Ostrogradsky:1850fid}, we can identify three canonical variables: 
\begin{equation}
    \Psi_1=\Psi_n,\qquad \Psi_2=\dot \Psi_n,\qquad\text{and}\qquad \chi.
\end{equation}
The corresponding canonical momentas are given by: 
\begin{equation}
    \pi_1=-8\beta\left(\frac{\partial^3\Psi_n}{\partial t^3}+\Delta\left(\dot \chi-\dot\Psi_n\right)\right),
\end{equation}
\begin{equation}
    \pi_2=8\beta\left(\Ddot{\Psi_n}+\Delta\left( \chi-\Psi_n\right)\right),
\end{equation}
and
\begin{equation}
    \pi_{\chi}=0. 
\end{equation}
As before, this theory has a primary constraint: 
\begin{equation}
    \pi_{\chi}\approx0,
\end{equation}
and the  Hamiltonian density is given by: 
\begin{equation}
    \mathcal{H}=\pi_1\Psi_2+\frac{1}{16\beta}\pi_2\pi_2+\pi_2\Delta\left(\Psi_1-\chi\right).
\end{equation}
Since we have brought the previous theory to a form described by the Hamiltonian for three canonical variables together with a constraint, we can straightforwardly apply the constraint analysis of \cite{Dirac}. The total Hamiltonian is given by: 
\begin{equation}
    \mathcal{H}_T=\pi_1\Psi_2+\frac{1}{16\beta}\pi_2\pi_2+\pi_2\Delta\left(\Psi_1-\chi\right)+v\pi_{\chi},
\end{equation}
where $v$ is the Lagrange multiplier. Similarly to the previous case, we can define the Poisson brackets: 
\begin{equation}
    \left\{f(\Vec{x},t),g(\Vec{y},t)\right\}=\sum_{i=1,2,3}\int d^3z\left( \frac{\delta f}{\delta Q_i(z)}\frac{\delta g}{\delta P_i(z)}-\frac{\delta g}{\delta Q_i(z)}\frac{\delta f}{\delta P_i(z)}\right),
\end{equation}
where 
\begin{equation}
    Q_i=\left\{\Psi_1,\Psi_2,\chi\right\},\qquad\text{and}\qquad  P_i=\left\{\pi_1,\pi_2,\pi_{\chi}\right\}. 
\end{equation}
Then, the equation of motion for the function $f$ is given by: 
\begin{equation}
    \dot f= \left\{f,H_T\right\}.
\end{equation}
From the consistency condition, 
\begin{equation}
    \dot\pi_{\chi}=0,
\end{equation}
we find the secondary constraint: 
\begin{equation}
    C_1=\Delta\pi_2=0\qquad\to\qquad \pi_2\approx0. 
\end{equation}
This one further implies another constraint: 
\begin{equation}
    \dot \pi_2=0\qquad\to\qquad C_2=-\pi_1. 
\end{equation}
Thus, in total, this theory has three first-class constraints: 
\begin{equation}
    \pi_{\chi}\approx0, \qquad\pi_1\approx0,\qquad \text{and}\qquad \pi_2\approx0. 
\end{equation}
The extended Hamiltonian is then given by: 
\begin{equation}
    H_E=\int d^3x\left(\pi_1\Psi_2+\frac{1}{16\beta}\pi_2\pi_2+\pi_2\Delta\left(\Psi_1-\chi\right)+v\pi_{\chi}+u_1\pi_1+u_2\pi_2\right),
\end{equation}
where $u_1$ and $u_2$ are Lagrange multipliers. The equations of motion for the canonical variables imply three \textit{dof}. Substracting the three constraints brings us to the overall zero \textit{dof} of this theory, in agreement witht the direct approach. 

\subsection*{\large\textsc{\textcolor{YellowOrange}{C} }\;\; \textsc{The limits of $R+R^2$ Gravity in the Einstein frame}}
\addcontentsline{toc}{subsection}{\protect\numberline{}\large\textsc{\textcolor{YellowOrange}{C} }\;\; \textsc{The limits of $R+R^2$ Gravity in the Einstein frame}}

 In this section, we will comment on the limits of the full $R+R^2$ gravity, described by the action (\ref{StarobinskyAction}) from the Einstein frame \cite{Ellis:2017xwz, Mukhanov:1989rq, Whitt:1984pd}. The action (\ref{StarobinskyAction}) can be requested in the following form: 
\begin{equation}
    S=\int d^4x\sqrt{-g}\left(M^2R+2\beta R\chi-\beta\chi^2\right). 
\end{equation}
Here, $\chi$ satisfies the constraint: 
\begin{equation}
    \chi=R. 
\end{equation}
By performing a conformal transformation of the metric: 
\begin{equation}
    g_{\mu\nu}\to\Tilde{g}_{\mu\nu}=\left(1+\frac{2\beta}{M^2}\chi\right)g_{\mu\nu},
\end{equation}
and upon defining the canonically normalized field:
\begin{equation}
    \chi_n=\sqrt{3}M\ln\left(1+\frac{2\beta}{M^2}\chi\right),
\end{equation}
the above action becomes: 
\begin{equation}\label{EinsteinactionStar}
    S=\int d^4x\sqrt{\Tilde{g}}\left[M^2\Tilde{R}-\frac{1}{2}\Tilde{g}^{\mu\nu}\partial_{\mu}\chi_n\partial_{\nu}\chi_n-\frac{M^4}{4\beta}\left(1-e^{-\frac{\chi_n}{\sqrt{3}M}}\right)^2\right],
\end{equation}
where $\Tilde{R}$ is the Ricci scalar corresponding to the new metric $\Tilde{g}_{\mu\nu}$. Let us now expand on the above potential, assuming that 
\begin{equation}\label{assumption}
    \frac{\chi_n}{M}<1. 
\end{equation}
In this case, the action can be approximated by: 
\begin{equation}
    S\sim\int d^4x\sqrt{\Tilde{g}}\left[M^2\Tilde{R}-\frac{1}{2}\Tilde{g}^{\mu\nu}\partial_{\mu}\chi_n\partial_{\nu}\chi_n-\frac{m^2_{\chi}}{2}\chi_n^2+\frac{M}{12\sqrt{3}\beta}\chi_n^3\right],
\end{equation}
where 
\begin{equation}
    m_{\chi}^2=\frac{M^2}{6\beta}
\end{equation}
is the mass of the scalar. Thus, we have found the action of a massive scalar, similar to the direct approach. 

It is clear that in the limit $M\to0$, both the tensor and the scalar mode will get strongly coupled, at the scale that we have previously found -- the tensor modes will obey the same behavior as in Einstein gravity, while the scalar will get strongly coupled at the same scale due to the interaction with the tensor mode. 

Let us now study the behavior of this scalar in the limit when $\beta\to0$, and for simplicity ignore the tensor perturbations. This limit is consistent only in the regime for scales 
\begin{equation}
    k^2\sim\frac{1}{L^2}\ll m_{\chi}^2. 
\end{equation}
As we have previously found, the minimal amplitude of quantum fluctuations for the scalar mode corresponding to these scales is given by
\begin{equation}
    \delta\chi_n\sim\frac{\beta^{1/4}}{(ML)^{3/2}}. 
\end{equation}
Let us now find the strong coupling scale, at which the non-linear terms become of the same order as linear ones on the level of the equation of motion. We can easily estimate this by working directly with the action. 
By estimating 
\begin{equation}
    \dot{\chi}_n\sim m_{\chi}\chi_n\qquad \text{and}\qquad \partial_i\chi_n\sim\frac{\chi_n}{L},
\end{equation}
we find: 
\begin{equation}
  \dot\chi_n\dot\chi_n\sim\frac{M}{\sqrt{\beta}L^3}\qquad\text{and}\qquad \frac{M}{\beta}\chi_n^3\sim\frac{M}{\beta^{1/4}L^3(ML)^{3/2}}. 
\end{equation}
The two terms are of the same order at length scales
\begin{equation}
    L_{\chi}\sim\frac{\beta^{1/6}}{M}
\end{equation}
which is precisely the scale that we have found with the direct approach. It is easy to verify that for scales $L>L_{\chi}$, the scalar field is weakly coupled. However, once the strong coupling scale is reached, the assumption (\ref{assumption}) will no longer hold. At this point, one has to work with the full potential, given in (\ref{EinsteinactionStar}). One could wonder if our estimate should be different, as the scalar field no longer has the mass term associated with it. However, it is easy to check that the potential term is dominating over the kinetic term at the strong coupling scale. Thus, at this point, the scalar is already strongly coupled -- it loses its linear propagator and becomes frozen due to the dominant potential. 

Finally, an interesting behavior happens in the limit when $\beta\to\infty$. In this case, the potential vanishes, while we are left with only Einstein gravity coupled to a scalar. However, at the same time, the conformal transformation is undefined in this case. Thus even though the pure $R^2$ theory has no dof, we could interpret the change in the description due to the singular behavior of the transformations that connect the two theories. It should be noted that in the case of the pure $R^2$ gravity \cite{Kehagias:2015ata} found non-trivial black-hole solutions with which the transformation between the conformal and Einstein frames is singular and thus which are not supported with both of the frames.

\subsection*{\large\textsc{\textcolor{YellowOrange}{D} }\;\; \textsc{Can we use Einstein and String frames for the Pure Theory?}}
\addcontentsline{toc}{subsection}{\protect\numberline{}\large\textsc{\textcolor{YellowOrange}{D} }\;\; \textsc{Can we use Einstein and String frames for the Pure Theory?}}
In this section, we will study if the action of the pure $R^2$ gravity can be studied in the Einstein or String frames. 

\begin{equation}\label{originalactionR2}
    S_{R^2}=\int d^4x \sqrt{-g}\beta R^2.
\end{equation}

\subsection*{\textsc{\textcolor{YellowOrange}{D.1} \;\;}\large\textsc{The Einstein Frame}}
\addcontentsline{toc}{subsubsection}{\protect\numberline{}\textsc{\textcolor{YellowOrange}{D.1} \;\;}\large\textsc{The Einstein Frame}}
Following the analysis of \cite{Alvarez-Gaume:2015rwa}, let us derive the corresponding action in the Einstein frame. For this, let us consider the following action:
\begin{equation}\label{conformalframeR2}
    S=\int d^4x \sqrt{-g}\left(\phi R-\frac{1}{4\beta}\phi^2\right).
\end{equation}
Here, $\phi$ satisfies the constraint: 
\begin{equation}
    \phi=2\beta R. 
\end{equation}
By substituting it into (\ref{conformalframeR2}), it can be easily shown that the action reduces to the original one, given by (\ref{originalactionR2}). Furthermore, let us define a conformally rescaled metric: 
\begin{equation}
    \Tilde{g}_{\mu\nu}=\frac{2}{M_{pl}^2}\phi g_{\mu\nu}
\end{equation}
By expressing (\ref{conformalframeR2}) in temrs of the new metric, we find: 
\begin{equation}\label{EisnteinFrameR2}
    S_{(E)}=\int d^4x\sqrt{-\Tilde{g}}\left[\frac{M_{pl}^2}{2}\left(\Tilde{R}-2\Lambda_{\beta}\right)-\frac{1}{2}\Tilde{g}^{\mu\nu}\partial_{\mu}\Phi_n\partial_{\nu}\Phi_n\right],
\end{equation}
where $\Tilde{R}$ is the curvature corresponding to the new metric,  
\begin{equation}
    \Lambda_{\beta}= \frac{M_{pl}^2}{16\beta},\qquad\text{and}\qquad \Phi_n=\sqrt{\frac{3}{2}}M_{pl}\ln\left(\frac{2\phi}{M_{pl}^2}\right)
\end{equation}
The action (\ref{EisnteinFrameR2}) is the action of the pure $R^2$ gravity in the Einstein frame. 

As pointed out in \cite{Alvarez-Gaume:2015rwa}, this formulation only works for $R\neq 0$. In the flat case, $R=0$, and thus we have $\phi=0$. As a result, the conformally rescaled metric $\Tilde{g}_{\mu\nu}$ is not defined, and we cannot use the Einstein frame in order to describe $R^2$ gravity in flat space-time. 

We could also intuitively understand why the connection between the original and Einstein frame is absent for the flat-space metric $g_{\mu\nu}=\eta_{\mu\nu}$ -- the action (\ref{EisnteinFrameR2}) has cosmological constant, and thus the flat space is not even the solution of the equations for the metric $\Tilde{g}_{\mu\nu}$. 

\subsection*{\textsc{\textcolor{YellowOrange}{D.2} \;\;}\large\textsc{The String Frame}}
\addcontentsline{toc}{subsubsection}{\protect\numberline{}\textsc{\textcolor{YellowOrange}{D.1} \;\;}\large\textsc{The String Frame}}
For non-zero curvature in the original (or conformal) frame, following the procedure of \cite{Gasperini:1993hu}, it can be shown that the string frame for the pure $R^2$ gravity is defined by:
\begin{equation}\label{stringframeR2}
    S_{(S)}=\frac{M_{pl}^2}{2}\int d^4x\sqrt{-g_S}e^{-\varphi}\left(R_S+g^{\mu\nu}_S\partial_{\mu}\varphi\partial_{\nu}\varphi-2\Lambda_{\beta}e^{-\varphi}\right).
\end{equation}

Here, $g_S$ and $R_S$ are the metric and Ricci scalar corresponding to the string-frame, while $\varphi$ is the scalar field. By conformally rescaling the above action: 
\begin{equation}
    g_{S\mu\nu}=e^{\varphi}\Tilde{g}_{\mu\nu}
\end{equation}
and by identifying 
\begin{equation}
    \Phi_n=\frac{M_{pl}}{\sqrt{2}}\varphi,
\end{equation}
it is easy to see that the action (\ref{stringframeR2}) becomes the action of the Einstein frame (\ref{EisnteinFrameR2}). 

Let us now consider the $R=0$ case. The string-frame metric is connected to the original one with: 
\begin{equation}
    g_{S\mu\nu}=\left(\frac{2\phi}{M_{pl}^2}\right)^{1+\sqrt{3}}g_{\mu\nu}. 
\end{equation}
Due to the constraint: 
\begin{equation}
    \phi=2\beta R,
\end{equation}
for $R=0$, we have $\phi=0$. Thus, the string-frame metric is undefined in this case, similarly to the Einstein-frame metric, $\Tilde{g}_{\mu\nu}$. 

Therefore, we have seen that the two frames are not well defined in the case of flat space. The way to study the perturbations of pure $R^2$ gravity is thus with the direct method, in the original frame.

\renewcommand\refname{\textcolor{Black}{\Large\textsc{\textbf{{References\hfill  }}}}}


\begin{thebibliography}{1}
\myfontsize

\bibitem{CANTATA:2021ktz}
E.~N.~Saridakis \textit{et al.} [CANTATA],
\textit{Modified Gravity and Cosmology: An Update by the CANTATA Network,}
Springer, 2021,
ISBN 978-3-030-83714-3, 978-3-030-83717-4, 978-3-030-83715-0
doi:10.1007/978-3-030-83715-0
[arXiv:2105.12582 [gr-qc]].

\bibitem{Utiyama:1962sn}
R.~Utiyama and B.~S.~DeWitt,
\textit{Renormalization of a classical gravitational field interacting with quantized matter fields,}
J. Math. Phys. \textbf{3} (1962), 608-618
doi:10.1063/1.1724264



\bibitem{Stelle:1976gc}
K.~S.~Stelle,
\textit{Renormalization of Higher Derivative Quantum Gravity,}
Phys. Rev. D \textbf{16} (1977), 953-969
doi:10.1103/PhysRevD.16.953

\bibitem{Stelle:1977ry}
K.~S.~Stelle,
\textit{Classical Gravity with Higher Derivatives,}
Gen. Rel. Grav. \textbf{9} (1978), 353-371
doi:10.1007/BF00760427

\bibitem{Boulware:1983td}
D.~G.~Boulware, G.~T.~Horowitz and A.~Strominger,
\textit{Zero Energy Theorem for Scale Invariant Gravity,}
Phys. Rev. Lett. \textbf{50} (1983), 1726
doi:10.1103/PhysRevLett.50.1726

\bibitem{Horowitz:1984wv}
G.~T.~Horowitz,
\textit{Quantum Cosmology With a Positive Definite Action,}
Phys. Rev. D \textbf{31} (1985), 1169
doi:10.1103/PhysRevD.31.1169

\bibitem{David:1984uv}
F.~David and A.~Strominger,
\textit{On the Calculability of Newton's Constant and the Renormalizability of Scale Invariant Quantum Gravity,}
Phys. Lett. B \textbf{143} (1984), 125-129
doi:10.1016/0370-2693(84)90817-7

\bibitem{Buchbinder:1987vp}
I.~L.~Buchbinder and S.~L.~Lyakhovich,
\textit{Canonical Quantization and Local Measure of R**2 Gravity,}
Class. Quant. Grav. \textbf{4} (1987), 1487-1501
doi:10.1088/0264-9381/4/6/008


\bibitem{Deser:2007vs}
S.~Deser and B.~Tekin,
\textit{New energy definition for higher curvature gravities,}
Phys. Rev. D \textbf{75} (2007), 084032
doi:10.1103/PhysRevD.75.084032
[arXiv:gr-qc/0701140 [gr-qc]].



\bibitem{tHooft:2011aa}
G.~'t Hooft,
\textit{A class of elementary particle models without any adjustable real parameters,}
Found. Phys. \textbf{41} (2011), 1829-1856
doi:10.1007/s10701-011-9586-8
[arXiv:1104.4543 [gr-qc]].

\bibitem{Lu:2011ks}
H.~Lu, Y.~Pang and C.~N.~Pope,
\textit{Conformal Gravity and Extensions of Critical Gravity,}
Phys. Rev. D \textbf{84} (2011), 064001
doi:10.1103/PhysRevD.84.064001
[arXiv:1106.4657 [hep-th]].

\bibitem{Park:2012ds}
M.~Park and L.~Sorbo,
\textit{Massive Gravity from Higher Derivative Gravity with Boundary Conditions,}
JHEP \textbf{01} (2013), 043
doi:10.1007/JHEP01(2013)043
[arXiv:1210.7733 [hep-th]].

\bibitem{Alvarez-Gaume:2015rwa}
L.~Alvarez-Gaume, A.~Kehagias, C.~Kounnas, D.~L\"ust and A.~Riotto,
\textit{Aspects of Quadratic Gravity,}
Fortsch. Phys. \textbf{64} (2016) no.2-3, 176-189
doi:10.1002/prop.201500100
[arXiv:1505.07657 [hep-th]].

\bibitem{Salvio:2018crh}
A.~Salvio,
\textit{Quadratic Gravity}
Front. in Phys. \textbf{6} (2018), 77
doi:10.3389/fphy.2018.00077
[arXiv:1804.09944 [hep-th]].

\bibitem{DeFelice:2023apt}
A.~De Felice, R.~Kawaguchi, K.~Mizui and S.~Tsujikawa,
\textit{Weyl Starobinsky inflation,}
[arXiv:2309.01835 [gr-qc]].

\bibitem{DeFelice:2023kpl}
A.~De Felice and S.~Tsujikawa,
\textit{Stability of Schwarzshild black holes in quadratic gravity with Weyl curvature domination,}
[arXiv:2307.06490 [gr-qc]].

\bibitem{Buoninfante:2023ryt}
L.~Buoninfante,
\textit{Massless and Partially Massless Limits in Quadratic Gravity,}
[arXiv:2308.11324 [hep-th]]. 


\bibitem{Tadros:2023teq}
P.~Tadros and I.~Kol\'a\v{r},
\textit{Carrollian limit of quadratic gravity,}
[arXiv:2307.13760 [gr-qc]].

\bibitem{Manolakos:2019fle}
G.~Manolakos, P.~Manousselis and G.~Zoupanos,
\textit{Four-dimensional Gravity on a Covariant Noncommutative Space,}
JHEP \textbf{08} (2020), 001
doi:10.1007/JHEP08(2020)001
[arXiv:1902.10922 [hep-th]].

\bibitem{Manolakos:2021rcl}
G.~Manolakos, P.~Manousselis and G.~Zoupanos,
\textit{Four-Dimensional Gravity on a Covariant Noncommutative Space (II),}
Fortsch. Phys. \textbf{69} (2021) no.8-9, 2100085
doi:10.1002/prop.202100085
[arXiv:2104.13746 [hep-th]].

\bibitem{Tekin:2016vli}
B.~Tekin,
\textit{Particle Content of Quadratic and $f(R_{\mu\nu\sigma \rho})$ Theories in $(A)dS$,}
Phys. Rev. D \textbf{93} (2016) no.10, 101502
doi:10.1103/PhysRevD.93.101502
[arXiv:1604.00891 [hep-th]].

\bibitem{Konitopoulos:2023wst}
S.~Konitopoulos, D.~Roumelioti and G.~Zoupanos,
\textit{Unification of Gravity and Internal Interactions,}
[arXiv:2309.15892 [hep-th]].

\bibitem{Fradkin:1981iu}
E.~S.~Fradkin and A.~A.~Tseytlin,
\textit{Renormalizable asymptotically free quantum theory of gravity,}
Nucl. Phys. B \textbf{201} (1982), 469-491
doi:10.1016/0550-3213(82)90444-8

\bibitem{Julve:1978xn}
J.~Julve and M.~Tonin,
\textit{Quantum Gravity with Higher Derivative Terms,}
Nuovo Cim. B \textbf{46} (1978), 137-152
doi:10.1007/BF02748637



\bibitem{Capper:1975ig}
D.~M.~Capper and M.~J.~Duff,
\textit{Conformal Anomalies and the Renormalizability Problem in Quantum Gravity,}
Phys. Lett. A \textbf{53} (1975), 361
doi:10.1016/0375-9601(75)90030-4

\bibitem{Ostrogradsky:1850fid}
M.~Ostrogradsky,
\textit{M\'emoires sur les \'equations diff\'erentielles, relatives au probl\`eme des isop\'erim\`etres,}
Mem. Acad. St. Petersbourg \textbf{6} (1850) no.4, 385-517

\bibitem{Riegert:1984hf}
R.~J.~Riegert,
\textit{THE PARTICLE CONTENT OF LINEARIZED CONFORMAL GRAVITY,}
Phys. Lett. A \textbf{105} (1984), 110-112
doi:10.1016/0375-9601(84)90648-0

\bibitem{Maldacena:2011mk}
J.~Maldacena,
\textit{Einstein Gravity from Conformal Gravity,}
[arXiv:1105.5632 [hep-th]].

\bibitem{Hell:2023rbf}
A.~Hell, D.~Lust and G.~Zoupanos,
\textit{On the ghost problem of conformal gravity,}
JHEP \textbf{08} (2023), 168
doi:10.1007/JHEP08(2023)168
[arXiv:2306.13714 [hep-th]].

\bibitem{Anastasiou:2016jix}
G.~Anastasiou and R.~Olea,
\textit{From conformal to Einstein Gravity,}
Phys. Rev. D \textbf{94} (2016) no.8, 086008
doi:10.1103/PhysRevD.94.086008
[arXiv:1608.07826 [hep-th]].

\bibitem{Anastasiou:2020mik}
G.~Anastasiou, I.~J.~Araya and R.~Olea,
\textit{Einstein Gravity from Conformal Gravity in 6D,}
JHEP \textbf{01} (2021), 134
doi:10.1007/JHEP01(2021)134
[arXiv:2010.15146 [hep-th]].

\bibitem{Sotiriou:2008rp}
T.~P.~Sotiriou and V.~Faraoni,
\textit{`f(R) Theories Of Gravity,}
Rev. Mod. Phys. \textbf{82} (2010), 451-497
doi:10.1103/RevModPhys.82.451
[arXiv:0805.1726 [gr-qc]].

\bibitem{Casado-Turrion:2023rni}
A.~Casado-Turri\'on, \'A.~de la Cruz-Dombriz and A.~Dobado,
\textit{Physical nonviability of a wide class of f(R) models and their constant-curvature solutions,}
Phys. Rev. D \textbf{108} (2023) no.6, 064006
doi:10.1103/PhysRevD.108.064006
[arXiv:2303.02103 [gr-qc]].

\bibitem{Mukhanov:1989rq}
V.~F.~Mukhanov,
\textit{Quantum Theory of Cosmological Perturbations in R(2) Gravity,}
Phys. Lett. B \textbf{218} (1989), 17-20
doi:10.1016/0370-2693(89)90467-X

\bibitem{Mukhanov1987}
L.~A.~Kofman, V.~F.~Mukhanov and D.~Y.~Pogosian,
\textit{Evolution of Inhomogeneities in Inflationary Models in a Theory of Gravitation With Higher Derivatives,}
Sov. Phys. JETP \textbf{66} (1987), 433-440








\bibitem{Starobinsky:1979ty}
A.~A.~Starobinsky,
\textit{Spectrum of relict gravitational radiation and the early state of the universe,}
JETP Lett. \textbf{30} (1979), 682-685

\bibitem{Planck:2018jri}
Y.~Akrami \textit{et al.} [Planck],
\textit{Planck 2018 results. X. Constraints on inflation,}
Astron. Astrophys. \textbf{641} (2020), A10
doi:10.1051/0004-6361/201833887
[arXiv:1807.06211 [astro-ph.CO]].

\bibitem{Mukhanov:1981xt}
V.~F.~Mukhanov and G.~V.~Chibisov,
\textit{Quantum Fluctuations and a Nonsingular Universe,}
JETP Lett. \textbf{33} (1981), 532-535

\bibitem{Guth:1980zm}
A.~H.~Guth,
\textit{The Inflationary Universe: A Possible Solution to the Horizon and Flatness Problems,}
Phys. Rev. D \textbf{23} (1981), 347-356
doi:10.1103/PhysRevD.23.347

\bibitem{Linde:1981mu}
A.~D.~Linde,
\textit{A New Inflationary Universe Scenario: A Possible Solution of the Horizon, Flatness, Homogeneity, Isotropy and Primordial Monopole Problems,}
Phys. Lett. B \textbf{108} (1982), 389-393
doi:10.1016/0370-2693(82)91219-9

\bibitem{Albrecht:1982wi}
A.~Albrecht and P.~J.~Steinhardt,
\textit{Cosmology for Grand Unified Theories with Radiatively Induced Symmetry Breaking,}
Phys. Rev. Lett. \textbf{48} (1982), 1220-1223
doi:10.1103/PhysRevLett.48.1220

\bibitem{Cecotti}
S. Cecotti, \textit{Higher Derivative Supergravity Is Equivalent To Standard Supergravity Coupled To
Matter. 1.},” Phys. Lett. B 190 (1987) 86.

\bibitem{Ellis:2013xoa}
J.~Ellis, D.~V.~Nanopoulos and K.~A.~Olive,
\textit{No-Scale Supergravity Realization of the Starobinsky Model of Inflation,}
Phys. Rev. Lett. \textbf{111} (2013), 111301
[erratum: Phys. Rev. Lett. \textbf{111} (2013) no.12, 129902]
doi:10.1103/PhysRevLett.111.111301
[arXiv:1305.1247 [hep-th]].

\bibitem{Farakos:2013cqa}
F.~Farakos, A.~Kehagias and A.~Riotto,
\textit{On the Starobinsky Model of Inflation from Supergravity,}
Nucl. Phys. B \textbf{876} (2013), 187-200
doi:10.1016/j.nuclphysb.2013.08.005
[arXiv:1307.1137 [hep-th]].

\bibitem{Ellis:2013nxa}
J.~Ellis, D.~V.~Nanopoulos and K.~A.~Olive,
\textit{Starobinsky-like Inflationary Models as Avatars of No-Scale Supergravity,}
JCAP \textbf{10} (2013), 009
doi:10.1088/1475-7516/2013/10/009
[arXiv:1307.3537 [hep-th]].

\bibitem{Kallosh:2013lkr}
R.~Kallosh and A.~Linde,
\textit{Superconformal generalizations of the Starobinsky model,}
JCAP \textbf{06} (2013), 028
doi:10.1088/1475-7516/2013/06/028
[arXiv:1306.3214 [hep-th]].

\bibitem{Ferrara:2013wka}
S.~Ferrara, R.~Kallosh and A.~Van Proeyen,
\textit{On the Supersymmetric Completion of $R+R^2$ Gravity and Cosmology,}
JHEP \textbf{11} (2013), 134
doi:10.1007/JHEP11(2013)134
[arXiv:1309.4052 [hep-th]].

\bibitem{Ferrara:2013pla}
S.~Ferrara, A.~Kehagias and M.~Porrati,
\textit{Vacuum structure in a chiral $\mathcal{R}+\mathcal{R}^n$ modification of pure supergravity,}
Phys. Lett. B \textbf{727} (2013), 314-318
doi:10.1016/j.physletb.2013.10.027
[arXiv:1310.0399 [hep-th]].

\bibitem{Ellis:2014cma}
J.~Ellis, N.~E.~Mavromatos and D.~V.~Nanopoulos,
\textit{Starobinsky-Like Inflation in Dilaton-Brane Cosmology,}
Phys. Lett. B \textbf{732} (2014), 380-384
doi:10.1016/j.physletb.2014.04.014
[arXiv:1402.5075 [hep-th]].

\bibitem{Ferrara:2014ima}
S.~Ferrara, A.~Kehagias and A.~Riotto,
\textit{The Imaginary Starobinsky Model,}
Fortsch. Phys. \textbf{62} (2014), 573-583
doi:10.1002/prop.201400018
[arXiv:1403.5531 [hep-th]].

\bibitem{Ferrara:2014fqa}
S.~Ferrara, A.~Kehagias and A.~Riotto,
\textit{The Imaginary Starobinsky Model and Higher Curvature Corrections,}
Fortsch. Phys. \textbf{63} (2015), 2-11
doi:10.1002/prop.201400070
[arXiv:1405.2353 [hep-th]].

\bibitem{Ferrara:2014yna}
S.~Ferrara and A.~Kehagias,
\textit{Higher Curvature Supergravity, Supersymmetry Breaking and Inflation,}
Subnucl. Ser. \textbf{52} (2017), 119-146
doi:10.1142/9789813148680\_0003
[arXiv:1407.5187 [hep-th]].

\bibitem{Ferrara:2014cca}
S.~Ferrara and M.~Porrati,
\textit{Minimal $R+R^2$ Supergravity Models of Inflation Coupled to Matter,}
Phys. Lett. B \textbf{737} (2014), 135-138
doi:10.1016/j.physletb.2014.08.050
[arXiv:1407.6164 [hep-th]].

\bibitem{Dalianis:2014aya}
I.~Dalianis, F.~Farakos, A.~Kehagias, A.~Riotto and R.~von Unge,
\textit{Supersymmetry Breaking and Inflation from Higher Curvature Supergravity,}
JHEP \textbf{01} (2015), 043
doi:10.1007/JHEP01(2015)043
[arXiv:1409.8299 [hep-th]].

\bibitem{Diamandis:2014vxa}
G.~A.~Diamandis, B.~C.~Georgalas, K.~Kaskavelis, P.~Kouroumalou, A.~B.~Lahanas and G.~Pavlopoulos,
\textit{Inflation in $R^2$ supergravity with non-minimal superpotentials,}
Phys. Lett. B \textbf{744} (2015), 74-81
doi:10.1016/j.physletb.2015.03.034
[arXiv:1411.5785 [hep-th]].

\bibitem{Lahanas:2015jwa}
A.~B.~Lahanas and K.~Tamvakis,
\textit{Inflation in no-scale supergravity,}
Phys. Rev. D \textbf{91} (2015) no.8, 085001
doi:10.1103/PhysRevD.91.085001
[arXiv:1501.06547 [hep-th]].

\bibitem{Kounnas:2014gda}
C.~Kounnas, D.~L\"ust and N.~Toumbas,
\textit{R$^2$ inflation from scale invariant supergravity and anomaly free superstrings with fluxes,}
Fortsch. Phys. \textbf{63} (2015), 12-35
doi:10.1002/prop.201400073
[arXiv:1409.7076 [hep-th]].

\bibitem{Farakos:2017mwd}
F.~Farakos, S.~Ferrara, A.~Kehagias and D.~L\"ust,
\textit{Non-linear Realizations and Higher Curvature Supergravity,}
Fortsch. Phys. \textbf{65} (2017) no.12, 1700073
doi:10.1002/prop.201700073
[arXiv:1707.06991 [hep-th]].

\bibitem{JAcobson}
T. Jacobson and R. C. Myers, \textit{Black hole entropy and higher curvature interactions,} Phys. Rev.
Lett. 70 (1993) 3684 [hep-th/9305016].

\bibitem{Frolov:2009qu}
V.~P.~Frolov and I.~L.~Shapiro,
\textit{Black Holes in Higher Dimensional Gravity Theory with Quadratic in Curvature Corrections,}
Phys. Rev. D \textbf{80} (2009), 044034
doi:10.1103/PhysRevD.80.044034
[arXiv:0907.1411 [gr-qc]].

\bibitem{Nelson:2010ig}
W.~Nelson,
\textit{Static Solutions for 4th order gravity,}
Phys. Rev. D \textbf{82} (2010), 104026
doi:10.1103/PhysRevD.82.104026
[arXiv:1010.3986 [gr-qc]].


\bibitem{Lu:2015cqa}
H.~Lu, A.~Perkins, C.~N.~Pope and K.~S.~Stelle,
\textit{Black Holes in Higher-Derivative Gravity,}
Phys. Rev. Lett. \textbf{114} (2015) no.17, 171601
doi:10.1103/PhysRevLett.114.171601
[arXiv:1502.01028 [hep-th]].

\bibitem{Edery:2014nha}
A.~Edery and Y.~Nakayama,
\textit{Restricted Weyl invariance in four-dimensional curved spacetime,}
Phys. Rev. D \textbf{90} (2014), 043007
doi:10.1103/PhysRevD.90.043007
[arXiv:1406.0060 [hep-th]].

\bibitem{Edery:2015wha}
A.~Edery and Y.~Nakayama,
\textit{Generating Einstein gravity, cosmological constant and Higgs mass from restricted Weyl invariance,}
Mod. Phys. Lett. A \textbf{30} (2015) no.30, 1550152
doi:10.1142/S0217732315501527
[arXiv:1502.05932 [hep-th]].

\bibitem{Oda:2020wdd}
I.~Oda,
\textit{Restricted Weyl symmetry,}
Phys. Rev. D \textbf{102} (2020) no.4, 045008
doi:10.1103/PhysRevD.102.045008
[arXiv:2005.04771 [hep-th]].


\bibitem{Kehagias:2015ata}
A.~Kehagias, C.~Kounnas, D.~L\"ust and A.~Riotto,
\textit{Black hole solutions in $R^{2}$ gravity,}
JHEP \textbf{05} (2015), 143
doi:10.1007/JHEP05(2015)143
[arXiv:1502.04192 [hep-th]].



\bibitem{Pravda:2016fue}
V.~Pravda, A.~Pravdova, J.~Podolsky and R.~Svarc,
\textit{Exact solutions to quadratic gravity,}
Phys. Rev. D \textbf{95} (2017) no.8, 084025
doi:10.1103/PhysRevD.95.084025
[arXiv:1606.02646 [gr-qc]].

\bibitem{Podolsky:2018pfe}
J.~Podolsky, R.~Svarc, V.~Pravda and A.~Pravdova,
\textit{Explicit black hole solutions in higher-derivative gravity,}
Phys. Rev. D \textbf{98} (2018) no.2, 021502
doi:10.1103/PhysRevD.98.021502
[arXiv:1806.08209 [gr-qc]].

\bibitem{Gurses:2012db}
M.~Gurses, T.~C.~Sisman and B.~Tekin,
\textit{New Exact Solutions of Quadratic Curvature Gravity,}
Phys. Rev. D \textbf{86} (2012), 024009
doi:10.1103/PhysRevD.86.024009
[arXiv:1204.2215 [hep-th]].


\bibitem{Nguyen:2023ufi}
H.~K.~Nguyen,
\textit{Non-triviality of asymptotically flat Buchdahl-inspired metrics in pure $R^2$ gravity,}
[arXiv:2305.12037 [gr-qc]].

\bibitem{Azreg-Ainou:2023qtf}
M.~Azreg-A\"\i{}nou and H.~K.~Nguyen,
\textit{A stationary axisymmetric vacuum solution for pure $R^2$ gravity,}
[arXiv:2304.08456 [gr-qc]].

\bibitem{Nguyen:2023kwr}
H.~K.~Nguyen and M.~Azreg-A\"\i{}nou,
\textit{Traversable Morris\textendash{}Thorne\textendash{}Buchdahl wormholes in quadratic gravity,}
Eur. Phys. J. C \textbf{83} (2023) no.7, 626
doi:10.1140/epjc/s10052-023-11805-3
[arXiv:2305.04321 [gr-qc]].

\bibitem{Nguyen:2022blj}
H.~K.~Nguyen,
\textit{Beyond Schwarzschild\textendash{}de Sitter spacetimes. II. An exact non-Schwarzschild metric in pure R2 gravity and new anomalous properties of R2 spacetimes,}
Phys. Rev. D \textbf{107} (2023) no.10, 104008
doi:10.1103/PhysRevD.107.104008
[arXiv:2211.03542 [gr-qc]].



\bibitem{Dent:2016efw}
J.~B.~Dent, D.~A.~Easson, T.~W.~Kephart and S.~C.~White,
\textit{Stability Aspects of Wormholes in $R^2$ Gravity,}
Int. J. Mod. Phys. D \textbf{26} (2017) no.10, 1750117
doi:10.1142/S0218271817501176
[arXiv:1608.00589 [gr-qc]].

\bibitem{Edery:2018jyp}
A.~Edery and Y.~Nakayama,
\textit{Gravitating magnetic monopole via the spontaneous symmetry breaking of pure $R^2$ gravity,}
Phys. Rev. D \textbf{98} (2018) no.6, 064011
doi:10.1103/PhysRevD.98.064011
[arXiv:1807.07004 [hep-th]].

\bibitem{Perapechka:2018iqo}
I.~Perapechka and Y.~Shnir,
\textit{$SU(2)$ Yang-Mills solitons in $R^2$ gravity,}
Phys. Lett. B \textbf{780} (2018), 152-158
doi:10.1016/j.physletb.2018.02.072
[arXiv:1801.07626 [hep-th]].


\bibitem{Duplessis:2015xva}
F.~Duplessis and D.~A.~Easson,
\textit{Traversable wormholes and non-singular black holes from the vacuum of quadratic gravity,}
Phys. Rev. D \textbf{92} (2015) no.4, 043516
doi:10.1103/PhysRevD.92.043516
[arXiv:1506.00988 [gr-qc]].

\bibitem{Bahamonde:2018zcq}
S.~Bahamonde, K.~Bamba and U.~Camci,
\textit{New Exact Spherically Symmetric Solutions in $f(R,\phi,X)$ gravity by Noether's symmetry approach,}
JCAP \textbf{02} (2019), 016
doi:10.1088/1475-7516/2019/02/016
[arXiv:1808.04328 [gr-qc]].

\bibitem{Whitt:1984pd}
B.~Whitt,
\textit{Fourth Order Gravity as General Relativity Plus Matter,}
Phys. Lett. B \textbf{145} (1984), 176-178
doi:10.1016/0370-2693(84)90332-0



\bibitem{Ruegg:2003ps}
H.~Ruegg and M.~Ruiz-Altaba,
\textit{The Stueckelberg field,}
Int. J. Mod. Phys. A \textbf{19} (2004), 3265-3348
doi:10.1142/S0217751X04019755
[arXiv:hep-th/0304245 [hep-th]].


\bibitem{Stueckelberg}
E. C. G. Stueckelberg, \textit{Die Wechselwirkungskräfte in der Elektrodynamik und in der Feldtheorie der Kräfte,} Helv. Phys. Acta 11, 225 1938.

\bibitem{Proca}
A. Proca, \textit{Sur la theorie ondulatoire des electrons positifs et negatifs}, J. Phys. Radium \textbf{7} (8), 347-353 (1936).

\bibitem{Hinterbichler:2011tt}
K.~Hinterbichler,
\textit{Theoretical Aspects of Massive Gravity,}
Rev. Mod. Phys. \textbf{84} (2012), 671-710
doi:10.1103/RevModPhys.84.671
[arXiv:1105.3735 [hep-th]].

\bibitem{deRham:2014zqa}
C.~de Rham,
\textit{Massive Gravity,}
Living Rev. Rel. \textbf{17} (2014), 7
doi:10.12942/lrr-2014-7
[arXiv:1401.4173 [hep-th]].



\bibitem{deRham:2011qq}
C.~de Rham, G.~Gabadadze and A.~J.~Tolley,
\textit{Helicity decomposition of ghost-free massive gravity,}
JHEP \textbf{11} (2011), 093
doi:10.1007/JHEP11(2011)093
[arXiv:1108.4521 [hep-th]].

\bibitem{Gambuti:2020onb}
G.~Gambuti and N.~Maggiore,
\textit{A note on harmonic gauge(s) in massive gravity,}
Phys. Lett. B \textbf{807} (2020), 135530
doi:10.1016/j.physletb.2020.135530
[arXiv:2006.04360 [gr-qc]].

\bibitem{Dvali:2007kt}
G.~Dvali, S.~Hofmann and J.~Khoury,
\textit{Degravitation of the cosmological constant and graviton width,}
Phys. Rev. D \textbf{76} (2007), 084006
doi:10.1103/PhysRevD.76.084006
[arXiv:hep-th/0703027 [hep-th]].


\bibitem{Kunimasa}
T. Kunimasa and T.Goto, \textit{Generalization of the Stueckelberg Formalism to the Massive Yang-Mills Field,} Prog. Theor. Phys. \textbf{37} (1967), 452-464.

\bibitem{Vainshtein}
A. I. Vainshtein and I. B. Khriplovich, \textit{On the zero-mass limit and renormalizability in the theory of massive yang-mills field,} Yad. Fiz. \textbf{13} (1971), 198-211. 

\bibitem{Huang:2007xf}
X.~Huang and L.~Parker,
\textit{Graviton Propagator in a Covariant Massive Gravity Theory,}
[arXiv:0705.1561 [hep-th]].


\bibitem{Ghosh:2023gvc}
J.~K.~Ghosh, E.~Kiritsis, F.~Nitti and V.~Nourry,
\textit{Quantum (in)stability of maximally symmetric space-times,}
[arXiv:2303.11091 [gr-qc]].

\bibitem{Edery:2023hxl}
A.~Edery,
\textit{Enlarging the symmetry of pure $R^2$ gravity, BRST invariance and its spontaneous breaking,}
[arXiv:2301.08638 [hep-th]].

\bibitem{Hinterbichler:2015soa}
K.~Hinterbichler and M.~Saravani,
\textit{St\"uckelberg approach to quadratic curvature gravity and its decoupling limits,}
Phys. Rev. D \textbf{93} (2016), 065006
doi:10.1103/PhysRevD.93.065006
[arXiv:1508.02401 [hep-th]].






\bibitem{Alvarez:2018lrg}
E.~Alvarez, J.~Anero, S.~Gonzalez-Martin and R.~Santos-Garcia,
\textit{Physical content of Quadratic Gravity,}
Eur. Phys. J. C \textbf{78} (2018) no.10, 794
doi:10.1140/epjc/s10052-018-6250-x
[arXiv:1802.05922 [hep-th]].



\bibitem{Kubo:2022jwu}
J.~Kubo and J.~Kuntz,
\textit{Spontaneous conformal symmetry breaking and quantum quadratic gravity,}
Phys. Rev. D \textbf{106} (2022) no.12, 126015
doi:10.1103/PhysRevD.106.126015
[arXiv:2208.12832 [hep-th]].

\bibitem{Kubo:2022dlx}
J.~Kubo, J.~Kuntz, J.~Rezacek and P.~Saake,
\textit{Inflation with massive spin-2 ghosts,}
JCAP \textbf{11} (2022), 049
doi:10.1088/1475-7516/2022/11/049
[arXiv:2207.14329 [astro-ph.CO]].

\bibitem{Kubo:2022lja}
J.~Kubo and J.~Kuntz,
\textit{Analysis of unitarity in conformal quantum gravity,}
Class. Quant. Grav. \textbf{39} (2022) no.17, 175010
doi:10.1088/1361-6382/ac8199
[arXiv:2202.08298 [hep-th]].



\bibitem{Kamimura:2021wzf}
A.~Kamimura and I.~Oda,
\textit{Quadratic gravity and restricted Weyl symmetry,}
Mod. Phys. Lett. A \textbf{36} (2021) no.19, 2150139
doi:10.1142/S021773232150139X
[arXiv:2103.11527 [hep-th]].

\bibitem{Dalianis:2020nuf}
I.~Dalianis, A.~Kehagias and I.~Taskas,
\textit{Higher Curvature Supergravity,}
PoS \textbf{CORFU2019} (2020), 152
doi:10.22323/1.376.0152

 \bibitem{Chibisov:1982nx}
G.~V.~Chibisov and V.~F.~Mukhanov,
\textit{Galaxy formation and phonons,}
Mon. Not. Roy. Astron. Soc. \textbf{200} (1982), 535-550

\bibitem{Mukhanov:1990me}
V.~F.~Mukhanov, H.~A.~Feldman and R.~H.~Brandenberger,
\textit{Theory of cosmological perturbations. Part 1. Classical perturbations. Part 2. Quantum theory of perturbations. Part 3. Extensions,}
Phys. Rept. \textbf{215} (1992), 203-333
doi:10.1016/0370-1573(92)90044-Z

\bibitem{Kodama:1984ziu}
H.~Kodama and M.~Sasaki,
\textit{Cosmological Perturbation Theory,}
Prog. Theor. Phys. Suppl. \textbf{78} (1984), 1-166
doi:10.1143/PTPS.78.1

\bibitem{Sasaki:1986hm}
M.~Sasaki,
\textit{Large Scale Quantum Fluctuations in the Inflationary Universe,}
Prog. Theor. Phys. \textbf{76} (1986), 1036
doi:10.1143/PTP.76.1036


\bibitem{Hell:2021wzm}
A.~Hell,
\textit{On the duality of massive Kalb-Ramond and Proca fields,}
JCAP \textbf{01} (2022) no.01, 056
doi:10.1088/1475-7516/2022/01/056
[arXiv:2109.05030 [hep-th]].

\bibitem{Chamseddine:2012gh}
A.~H.~Chamseddine and V.~Mukhanov,
\textit{Massive Hermitian Gravity,}
JHEP \textbf{08} (2012), 036
doi:10.1007/JHEP08(2012)036
[arXiv:1205.5828 [hep-th]].

\bibitem{Hell:2021oea}
A.~Hell,
\textit{The strong couplings of massive Yang-Mills theory,}
JHEP \textbf{03} (2022), 167
doi:10.1007/JHEP03(2022)167
[arXiv:2111.00017 [hep-th]].


 
\bibitem{Kawai}
H. Kawai,\textit{A Dual Transformation of the Nielsen-Olesen Model}, Prog. Theor. Phys. \textbf{65}, 351 (1981).

\bibitem{Trugenberger}
F. Quevedo and C. A. Trugenberger, \text{Phases of antisymmetric tensor field theories},
Nucl. Phys. B \textbf{501}, 501, 143-172 (1997).

\bibitem{Quevedo}
F. Quevedo, \textit{Duality and global symmetries}, Nucl. Phys. B Proc. Suppl. \textbf{61}, 23-41 (1998).

\bibitem{2001Smailagic}
A. Smailagic and E. Spallucci, \textit{The Dual phases of massless / massive Kalb-Ramond fields: Letter to the editor} J. Phys. A \textbf{34}  L435-L440 (2001).

\bibitem{2002Casini}
H. Casini, R. Montemayor and L. F. Urrutia, \textit{Duality for symmetric second rank tensors: The massive case}, Phys. Rev. D \textbf{66}, 085018 (2002).

\bibitem{Auria}
R. D'Auria and S. Ferrara, \textit{Dyonic masses from conformal field strengths in D even dimensions}, Phys. Lett. B \textbf{606}, 211-217 (2005).

\bibitem{Buchbinder}
I. L. Buchbinder, E. N. Kirillova, and N. G. Pletnev, \textit{Quantum Equivalence of Massive Antisymmetric Tensor Field Models in Curved Space}, Phys. Rev. D \textbf{78}, 084024 (2008).


\bibitem{Dalmazi}
D. Dalmazi and R. C. Santos, \textit{Spin-1 duality in $D$-dimensions}, Phys. Rev. D \textbf{84}, 045027 (2011).

\bibitem{Shifman}
M. Shifman and A. Yung, \textit{Hadrons of $\mathcal N=2$ Supersymmetric QCD in Four Dimensions from Little String Theory}, Phys. Rev. D \textbf{98}, no.8, 085013 (2018). 

\bibitem{Garcia}
G. B. De Gracia, \textit{Spin jumping in the context of a QCD effective model}, Int. J. Mod. Phys. A \textbf{32}, no.06n07, 1750041 (2017).

\bibitem{Kuzenko}
S. M. Kuzenko and K. Turner, \textit{Effective actions for dual massive (super) $p$-forms}, JHEP \textbf{01}, 040 (2021).

\bibitem{CF1980}
T. L. Curtright and P. G. O. Freund, \textit{MASSIVE DUAL FIELDS}, Nucl. Phys. B \textbf{172}, 413-424 (1980). 

\bibitem{C2019}
 T. L. Curtright, \textit{Massive dual spinless fields revisited}, Nucl. Phys. B \textbf{948}, 114784 (2019).

 \bibitem{Dvali}
 G. Dvali, \textit{Three-form gauging of axion symmetries and gravity}, [arXiv:hep-th/0507215 [hep-th]] (2005).


\bibitem{Fierz:1939ix}
M.~Fierz and W.~Pauli,
\textit{On relativistic wave equations for particles of arbitrary spin in an electromagnetic field,}
Proc. Roy. Soc. Lond. A \textbf{173} (1939), 211-232
doi:10.1098/rspa.1939.0140

\bibitem{vanDam:1970vg}
H.~van Dam and M.~J.~G.~Veltman,
\textit{Massive and massless Yang-Mills and gravitational fields,}
Nucl. Phys. B \textbf{22} (1970), 397-411
doi:10.1016/0550-3213(70)90416-5

\bibitem{Zakharov:1970cc}
V.~I.~Zakharov,
\textit{Linearized gravitation theory and the graviton mass,}
JETP Lett. \textbf{12} (1970), 312

\bibitem{Veltman:1968ki}
M.~J.~G.~Veltman,
\textit{Perturbation theory of massive Yang-Mills fields,}
Nucl. Phys. B \textbf{7} (1968), 637-650
doi:10.1016/0550-3213(68)90197-1


\bibitem{Reiff:1969pq}
J.~Reiff and M.~J.~G.~Veltman,
\textit{Massive yang-mills fields,}
Nucl. Phys. B \textbf{13} (1969), 545-564
doi:10.1016/0550-3213(69)90190-4

\bibitem{Slavnov:1972qb}
A.~A.~Slavnov,
\textit{Massive gauge fields,}
Teor. Mat. Fiz. \textbf{10} (1972), 305-328

\bibitem{Wong:1971er}
S.~K.~Wong,
\textit{Massless limit of the massive yang-mills field,}
Phys. Rev. D \textbf{3} (1971), 945-952
[erratum: Phys. Rev. D \textbf{3} (1971), 3243-3243]
doi:10.1103/PhysRevD.3.945

\bibitem{Boulware:1970zc}
D.~G.~Boulware,
\textit{Renormalizeability of massive non-abelian gauge fields - a functional integral approach,}
Annals Phys. \textbf{56} (1970), 140-171
doi:10.1016/0003-4916(70)90008-4
 
\bibitem{Vainshtein:1972sx}
A.~I.~Vainshtein,
\textit{To the problem of nonvanishing gravitation mass,}
Phys. Lett. B \textbf{39} (1972), 393-394
doi:10.1016/0370-2693(72)90147-5

\bibitem{Deffayet:2001uk}
C.~Deffayet, G.~R.~Dvali, G.~Gabadadze and A.~I.~Vainshtein,
\textit{Nonperturbative continuity in graviton mass versus perturbative discontinuity,}
Phys. Rev. D \textbf{65} (2002), 044026
doi:10.1103/PhysRevD.65.044026
[arXiv:hep-th/0106001 [hep-th]].

\bibitem{Gruzinov:2001hp}
A.~Gruzinov,
\textit{On the graviton mass,}
New Astron. \textbf{10} (2005), 311-314
doi:10.1016/j.newast.2004.12.001
[arXiv:astro-ph/0112246 [astro-ph]].

\bibitem{Hu:2023xcf}
Y.~M.~Hu, Y.~Yu, Y.~F.~Cai and X.~Gao,
\textit{The effective field theory approach to the strong coupling issue in $f(T)$ gravity with a non-minimally coupled scalar field,}
[arXiv:2311.12645 [gr-qc]].

\bibitem{Hu:2023juh}
Y.~M.~Hu, Y.~Zhao, X.~Ren, B.~Wang, E.~N.~Saridakis and Y.~F.~Cai,
\textit{The effective field theory approach to the strong coupling issue in f(T) gravity,}
JCAP \textbf{07} (2023), 060
doi:10.1088/1475-7516/2023/07/060
[arXiv:2302.03545 [gr-qc]].



\bibitem{Bardeen:1980kt}
J.~M.~Bardeen,
\textit{Gauge Invariant Cosmological Perturbations,}
Phys. Rev. D \textbf{22} (1980), 1882-1905
doi:10.1103/PhysRevD.22.1882

\bibitem{Cosmo}
 V. Mukhanov, \textit{Physical Foundations of Cosmology}, Cambridge University Press (2005).


\bibitem{Chamseddine:2018gqh}
A.~H.~Chamseddine and V.~Mukhanov,
\textit{Mimetic Massive Gravity: Beyond Linear Approximation,}
JHEP \textbf{06} (2018), 062
doi:10.1007/JHEP06(2018)062
[arXiv:1805.06598 [hep-th]].


\bibitem{QFTCS}
 V. Mukhanov, S. Winitzki, \textit{Introduction to Quantum Effects in Gravity}, Cambridge University Press (2007).


\bibitem{Dirac}
P.A.M. Dirac, \textit{Lectures on Quantum Mechanics}, Belfer Graduate School Monograph
Series, No.2 Yeshiva University, New York, (1964).

\bibitem{Woodard:2015zca}
R.~P.~Woodard,
\textit{Ostrogradsky's theorem on Hamiltonian instability,}
Scholarpedia \textbf{10} (2015) no.8, 32243
doi:10.4249/scholarpedia.32243
[arXiv:1506.02210 [hep-th]].

\bibitem{Ellis:2017xwz}
J.~Ellis, D.~V.~Nanopoulos and K.~A.~Olive,
\textit{From $R^2$ gravity to no-scale supergravity,}
Phys. Rev. D \textbf{97} (2018) no.4, 043530
doi:10.1103/PhysRevD.97.043530
[arXiv:1711.11051 [hep-th]].


\bibitem{Gasperini:1993hu}
M.~Gasperini and G.~Veneziano,
\textit{Inflation, deflation, and frame independence in string cosmology,}
Mod. Phys. Lett. A \textbf{8} (1993), 3701-3714
doi:10.1142/S0217732393003433
[arXiv:hep-th/9309023 [hep-th]].

\end{thebibliography}
\end{document}